\documentclass[12pt]{iopart}

\usepackage{graphicx}
\usepackage{amsmath}
\usepackage{amssymb}
\usepackage{color}
\usepackage{caption}
\usepackage{subcaption}

\newcommand{\blue}[1]{\textcolor{black}{#1}}
\newcommand{\dd}{\text{d}}

\begin{document}

\title[$\Phi_{1}$ and its impact on impurity transport]{Electrostatic potential variation on the flux surface and its impact on impurity transport}

\author{J. M. Garc\'ia-Rega\~na$^{1,2}$, C. D. Beidler$^1$, Y. Turkin$^1$, R. Kleiber$^1$, P. Helander$^1$, H. Maa\ss berg$^1$, J. A. Alonso$^3$ and J. L. Velasco$^3$}

\address{$^1$ Max-Planck-Institut f\"ur Plasmaphysik, 
EURATOM Association, Boltzmannstr. 2, 85748 Garching,Germany\\
$^2$ Max-Planck-Institut f\"ur Plasmaphysik, 
EURATOM Association, Wendelsteinstr. 1, 17491 Greifswald, Germany\\
$^3$ Laboratorio Nacional de Fusi\'on, Av. Complutense 40, 28040 Madrid, Spain}

\ead{jose.regana@ipp.mpg.de}

\begin{abstract}
The particle transport of impurities in magnetically confined 
plasmas under some conditions
does not find, neither quantitatively nor qualitatively,
a satisfactory theory-based explanation. This compromise 
the successful realization of thermo-nuclear fusion for energy production 
since its accumulation is known to be one of the causes that leads
to the plasma breakdown.\\
In standard reactor-relevant conditions 
this accumulation is in most stellarators intrinsic 
to the lack of toroidal symmetry, that leads to the 
neoclassical electric field to point radially inwards. 
This statement, that 
the standard theory allows to formulate, has been 
contradicted by some experiments that showed weaker or no accumulation
under such conditions \cite{Ida_pop_16_056111_2009, Yoshinuma_nf_49_062002_2009}.\\
The charge state of the impurities makes its transport more sensitive to the electric fields.
Thus, the short length scale turbulent electrostatic potential or 
its long wave-length variation on the flux surface $\Phi_{1}$ -- that
the standard neoclassical approach usually neglects --
might possibly shed some light on the 
experimental findings. In the present work the focus is put
on the second of the two, and investigate its influence of 
the radial transport of C$^{6+}$. We show that in LHD it is 
strongly modified by $\Phi_{1}$, both resulting in mitigated/enhanced accumulation at
 internal/external radial positions; for Wendelstein 7-X, on the contrary, $\Phi_{1}$ 
is expected to be considerably smaller and the transport
of C$^{6+}$ not affected up to an appreciable extent; and in TJ-II
the potential shows a moderate impact despite of the large amplitude
of $\Phi_1$ for the parameters considered.

\end{abstract}

\section{Introduction}

From a passive viewpoint the presence of impurities in thermo-nuclear fusion plasmas 
is intrinsic and is bound to the main wall material, whose 
choice in turn depends on the desired wall recycling properties.
Furthermore the deliberate introduction of certain impurity species 
is widely recognized as a powerful technique for the reduction 
of the heat exhaust on the plasma facing components to an tolerable level (see e.g. 
\cite{Kallenbach_ppcf_55_124041_2013} and reference therein).
The high radiative efficiency of impurities, exploited for that purpose,
results at the same time in a disadvantage if they
accumulate in the plasma core -- that apart from diluting the fusion fuel -- 
in some cases can lead to the radiative collapse of the plasma. 
The balance between 
the required concentration in the edge region and the 
lack of its accumulation in the core undoubtedly needs of the identification of
the physical mechanisms that determine the radial
particle flux of impurities, leading them to accumulate
or to avoid or mitigate that accumulation.\\
To that respect, in stellarators and heliotrons the standard physical picture
can be summarized as follows. The lack of toroidal symmetry of 
the magnetic field structure in these devices leads
to the existence of a population of particles trapped in the 
helical wells that determines the fluxes. Unless
that the magnetic field is such that the net 
radial drift of these particles is sufficiently reduced, the total flux surface average
radial current does not vanish.
In standard conditions where the bulk ions and electrons with masses
$m_{i}$ and $m_{e}$ have similar temperature $T_{i}\sim T_{e}$
their particle diffusion coefficients
are such that $D_{i}\sim\sqrt{m_{i}/m_{e}} D_{e}$ \cite{Ho_pf_30.2_1987}.
The system needs then of a radial electric field, so-called \textit{ambipolar},
that reduces the ion radial particle flux to the order of that for
the electrons.
This reduction is driven by the confining role that 
the $E\times B$ poloidal precession related to it has on the 
bulk ions. Tipically the electron confinement is not affected to the same extent 
due to the more frequent collisions they undergo ($\nu_{e}\sim\sqrt{m_{i}/m_{e}}\nu_{i}$), which prevent them from completing 
their poloidal precession orbits.\\
Writing the ambipolar electric field as 
$\mathbf{E}_{r}=E_{r}\nabla r$, with $E_{r}=-\dd \Phi_{0}/\dd r$ and 
$\Phi_{0}=\Phi_{0}(r)$ the electrostatic potential depending by definition only on the flux
surface label $r$, it is generally the case that under the previous conditions $E_{r}<0$ (\textit{ion root regime}), 
or in other words that the ambipolar electric field points radially inwards.
The importance of this for the accumulation of impurities is clear
on the context of the standard neoclassical theory.\\

The particle flux density across the flux surfaces
for the species $\alpha$ can be expressed in terms of the
neoclassical transport coefficients $L_{ij}^{\alpha}$
and the thermodynamic forces as follows, see
e.g. ref. \cite{Beidler_nf_51_076001_2011},

\begin{equation}
\label{pflux}
\left<\boldsymbol{\Gamma}_{\alpha}\cdot\nabla r\right>=-n_{s}L_{11}^{s}
\left[\frac{n'_{\alpha}}{n_{\alpha}}-\frac{Z_{\alpha}eE_{r}}{T_{\alpha}}+
\left(\frac{L_{12}^{\alpha}}{L_{11}^{\alpha}}-\frac{3}{2}\right)\frac{T'_{\alpha}}{T_{\alpha}}\right].
\end{equation}

In eq.~(\ref{pflux}), $\alpha$ is the species index, 
$n$ is the particle density, $T$ the temperature, $q$ and $Z$ 
are respectively the charge and charge state of the species, $e$ is the unit charge modulus and
the prime represents the derivative respect to $r$.
Therefore it is straightforward to conclude that $E_{r}<0$ drives particles towards
the center of the plasma and out of it otherwise. The weight
of the charge of the species $Z$ leads inevitably to 
the dominance of the term driving transport through $E_{r}$ as the 
$Z$ increases, and subsequently 
supports a stronger accumulation of impurities 
as the ionization state of these is higher.
In a tokamak the accumulation can be moderated by 
the so-called \textit{temperature screening}.
This mechanism rests on the fact that 
the term preceding the temperature gradient in eq.~(\ref{pflux}) can 
be negative for ions in the long mean free path regime (\textit{lmfp}).
Subsequently for peaked temperature
profiles the inwards flux of bulk ions driven by $T'$ enforces
the outwards flux of impurities in order to preserve quasi-neutrality
along the radial direction.
Nevertheless this mechanism cannot take place in stellarators
since the asymptotic scaling of the transport coefficients with the collisionality, 
inversely proportional to $\nu$ and $\sqrt{\nu}$ in the \textit{lmfp}, is such that
$L_{12}^{\alpha}/L_{11}^{\alpha}- 3/2 >0$ in both cases.
Thus the standard prediction for the radial particle transport
of impurities in non-axisymmetric devices is roughly speaking of accumulation in 
ion root regime and absence of it in electron root plasmas \cite{Maassberg_ppcf_41_1999}
where $E_{r}>0$.\\
Nevertheless, the standard neoclassical picture considering only the ambipolar
electrostatic potential $\Phi_0$ as constant on each flux surface
is particularly questionable in non-axisymmetric systems \blue{if
impurities of moderate to high charge are considered. This is due 
to two reasons: 1) regarding the approximation of the full electrostatic potential 
to only the ambipolar part $\Phi_0$ dismisses a potential variation on the flux surface 
$\Phi_1$ that in non-axisymmetric system can be particularly large. This is due to
the different dependency of $f_1$ (first order departure of the distribution function from its equilibrium $f_0$) 
with the collision frequency than that given in axi-symmetric system, and subsequently the perturbed linear 
density $n_1$ and related charge density and potential; 2) the fact that the weight of the $E\times B$ drift and
streaming acceleration related to $\Phi_{1}$ respect
to the magnetic drift and mirror acceleration scale proportionally to 
the charge state $Z$.}\\

\blue{Regarding the first of the aforementioned causes, 
in an axi-symmetric tokamak the effect of the radial electric field 
is weak and the $E_{r}\times B$ drift is negligible at zeroth order in
the normalized Larmor radius to the system size drift kinetic 
expansion parameter: $\rho_{*}=\rho/L$. Considering a Fourier-Legendre expansion of $f_1$ 
-- Fourier in the poloidal coordinate $\theta$ 
and Legendre in the pitch angle variable $\xi=v_{\|}/v$ --
the first order drift kinetic equation (DKE) 
shows that the structure of $f_1$ is such that the only non-zero Legendre coefficients 
are the even ones for the $\cos(m\theta)$
terms and the odd for the $\sin(m\theta)$ terms. The latter results in addition 
proportional to the collision frequency. Thus the only terms that can 
drive a density, and subsequently potential variation on the flux surface
formally tend to zero in the limit of vanishing collisionality
and will have a $\sin(m\theta)$ components only. Here $m$ is the Fourier
poloidal mode number, $v_{\|}$ is the parallel velocity and $v$
is the velocity modulus.\\
On the other hand the requirement of keeping the $E_{r}\times B$ drift 
at lowest order particle trajectory in non-axisymmetric 
systems breaks that structure in $f_1$. In this case the 
bounce averaged solution of the DKE $\left<f_1\right>_{b}$
results in a $\cos\theta$ structure with amplitude scaling with $v_{d}/\Omega_E$. 
With $v_{d}$ the modulus of the magnetic drift velocity and $\Omega_{E}=E_{r}/rB_0$
the poloidal $E_{r}\times B$ precession frequency. This picture 
where a finite non-vanishing variation is expected finds a cartoon in the fact
that in non-axisymmetric system the helically trapped particles shift their poloidal precession orbits
from the \textit{birth} flux surface due to the action of both the magnetic drift $\mathbf{v}_{d}$ and 
the poloidal $E_{r}\times B$ drift $\mathbf{v}_{E0}$ \cite{Ho_pf_30.2_1987}}. These 
drift velocities are expressed by

\begin{align}
&\mathbf{v}_{E0}=-\frac{\nabla\Phi_{0}\times\mathbf{B}}{B^{2}}
\label{eq:ve0},\\
&\mathbf{v}_{\text{d}}=\frac{m}{q}\frac{v_{\|}^{2}+\mu B}{B^{2}}
\mathbf{b}\times\nabla B,
\label{eq:vd}
\end{align}

with $\mathbf{B}$ the magnetic field vector, $B$ its modulus and $\mathbf{b}\equiv\mathbf{B}/B$,
$v_{\|}$ and $v_{\bot}$ the parallel and perpendicular components of the velocity and
$\mu=v_{\bot}^{2}/2B$ the magnetic moment.\\

\blue{
The second of the reasons aforementioned about 
the possible impact of $\Phi_1$ on impurities, follows from the 
comparison between $\Phi_1$  and $B$ as sources of transport and trapping.
Considering the $E\times B$ drift related to $\Phi_{1}$}

\begin{equation}
\mathbf{v}_{E1}=-\frac{\nabla\Phi_{1}\times \mathbf{B}}{B^{2}},
\end{equation}

it is in comparison to $\mathbf{v}_{\text{d}}$ is of the following order,

\begin{equation}
\frac{v_{E1}}{v_{\text{d}}}=\frac{Ze\Phi_{1}}{T}\frac{R}{a},
\label{eq:ve1_over_vd}
\end{equation}

where it has been assumed that $\Phi_{1}$ and the toroidal 
component of $B$ have typical variation length-scales 
$\nabla B \sim R^{-1}B$ and $\nabla \Phi_{1} \sim a^{-1}\Phi_{1}$.
The ratio shown in eq. \ref{eq:ve1_over_vd} indicates 
that even for low values of the ratio  $e\Phi_{1}/T$, $\Phi_{1}$
can be a source of impurities transport of the same magnitude than the 
magnetic field gradient and curvature.\\
Moreover, the streaming acceleration related to $\Phi_{1}$,
$a_{\text{s}}=-Ze\mathbf{b}\cdot\nabla\Phi_{1}$ and the magnetic mirror 
$a_{\text{m}}=-\mu\mathbf{b}\cdot\nabla B$ inside the helical wells 
are of order

\begin{equation}
\label{eq:as_over_am}
\frac{a_{\text{s}}}{a_{\text{m}}}=\frac{Ze\Phi_{1}}{T}\frac{B}{\Delta B},
\end{equation}

where $\Delta B$ is the amplitude of the helical ripple, and it has
been assumed that the variation length-scale of the helical component
of $B$ is of order $a$.
Then, the boundaries of the trapping regions considering $\Phi_{1}$
for the case of impurities can substantially be modified by those 
determined by only the magnetic field structure.\\

The impact of $\Phi_{1}$ on the transport has been studied in the past both 
analytically and numerically \cite{Mynick_pf_27.8_1984, Ho_pf_30.2_1987, Beidler_isw_2005}.
The moderate impact found on the 
bulk species has motivated the neglect of $\Phi_{1}$ in the standard approach for 
studying their neoclassical transport, but its neglect on impurity transport studies has not been fully justified
in non-axisymmetric system or studies yet apart from recent works \cite{Regana_ppcf_55_074008_2013}. 
The arguments presented before 
and expressed in the ratios \ref{eq:ve1_over_vd} and \ref{eq:as_over_am} motivates this work, 
where the radial particle transport of C$^{6+}$ impurities 
is studied including $\Phi_{1}$. 
The calculations including $\Phi_{1}$ requires relaxing
the usual mono-energetic assumption, and to that end they have 
been performed with the Monte Carlo code EUTERPE for an LHD like 
equilibrium, one standard configuration of the Wendelstein 7-X (W7-X) 
stellarator and a standard configuration of TJ-II. The main 
features of the code are described in sec. \ref{sec:euterpe}
and the results are shown in sec. \ref{sec:results}. 
Finally a summary on the amplitude of $\Phi_1$ given in these 
three devices and the conclusions are presented respectively in sections \ref{sec:remarks} \ref{sec:summary}.\\

\section{The code EUTERPE and the calculation of $\Phi_{1}$}
\label{sec:euterpe}

The calculations of the radial fluxes for the impurities and $\Phi_{1}$ 
has been carried out using the Monte Carlo
$f_1$ particle in cell (PIC) code EUTERPE \cite{Kornilov_nf_45.4_2005}. The current version 
in the gyro-kinetic modality is electro-magnetic, non-linear and 
considers the full surface and radial domain.
In the neoclassical version applied to the present problem the transport is assumed radially local instead.
It performs the collision of the $\delta f$ (that represents $f_{1}$ here)
off the equilibrium distribution function $f_{0}$,
applying at every time step a random 
change in the pitch angle of each marker after the integration of
the collisionless trajectory \cite{Takizuca_jcp_25.3_1977, Kauffmann_jpcs_260.1_2010}. 
The collision frequency for the colliding species $a$ is set as the sum of the 
deflection collision frequency of $a$ over 
all the target species $b$, including $a$ itself: $\nu_{a}=\sum_{ab}\nu^{\text{D}}_{ab}$.
The calculation of $E_{r}$ under the local ansatz is bound to be
carried out by iterative adjustment of the value of $E_{r}$ until the 
ambipolarity of the fluxes is fulfilled. Since
the computational cost is related to the time the fluxes take until 
their convergence and the number of iterations, 
for the calculations presented in sec. \ref{sec:results},
$E_{r}$ is a precalculated input obtained with a transport code \cite{Turkin_pop_18_022505_2011} based 
on DKES \cite{Hirshman_pf_29_2951_1986, Rij_pfb_1_563_1989}
or with the neoclassical code GSRAKE \cite{Beidler_ppcf_43_2001}.\\

The local \textit{ansatz} considers that the drifts across the flux
surfaces -- in this case the magnetic drifts
$\mathbf{v}_{\text{d}}$ and $E\times B$ drift $\mathbf{v}_{E1}$ related to $\Phi_{1}$ --
are in modulus much smaller -- of order $\rho_{*}$ -- respect to the parallel velocity, that in 
turn is of order of the thermal velocity
$v_{\text{th}}$. The same assumption is considered for the $f_1$ part of the distribution 
function respect to the equilibrium $f_{0}$. 
This justifies the neglect of the second order terms in the left hand side of the kinetic equation 
$\mathbf{v}_{\text{d}}\cdot\nabla f_1$ and $\mathbf{v}_{E1}\cdot\nabla f_1$.
Since the only drift retained at lowest order is the $E\times B$ drift
related to the ambipolar electric field $\mathbf{v}_{E0}$, and this pushes the markers
within the flux surface and not across it, each flux surface can be loaded with markers and considered separately
of the other integrating in this local limit the following characteristic trajectories in the 
phase space $(\mathbf{R},v_{\|}, \mu)$:

\begin{eqnarray}
&&\dot{\mathbf{R}}=v_{\|}\mathbf{b}-\frac{\nabla\Phi_{0}\times\mathbf{B}}{B^2}
\label{eq:dotR}\\
&&\dot{v}_{\|}=-\frac{Ze}{m}\mathbf{b}\cdot\nabla\Phi_{1}-\mu\mathbf{b}\cdot\nabla B-
\frac{v_{\|}}{B^{2}}\left(\mathbf{b}\times\nabla B\right)\cdot\nabla\Phi_{0},
\label{eq:dotvpar}\\
&&\dot{\mu}=0
\label{eq:dotmu}.
\end{eqnarray}

\blue{followed, as aforementioned, by a random kick on the pitch angle of each marker 
that performs the collisional step. Then the Vlasov equation integrated has the following form, 
\begin{equation}
\label{eq:vlasov}
\frac{\partial f_1}{\partial t}+
\dot{\mathbf{R}}\cdot\nabla f_1+
\dot{v}_{\|}\frac{\partial f_1}{\partial v_{\|}}=
-f_{\text{M}}\left(\mathbf{v}_{\text{d}}+\mathbf{v}_{E1}\right)\cdot\nabla r
\left[\frac{n'}{n}+\frac{q}{T}\Phi_0'+
\left(\frac{mv^{2}}{2T}-\frac{3}{2}+\frac{q}{T}\Phi_1\right)\frac{T'}{T}\right]
\end{equation}
where $f_{M}=\left[n_{0}/(2\pi)^{3/2} v_{\text{th}}^{3}\right]\exp\left[-\left(v_{\|}^{2}+v_{\bot}^2\right)/2v_{\text{th}}^{2}\right]$ is the standard local Maxwellian. To obtain eq. \ref{eq:vlasov} 
the equilibrium distribution function includes the Boltzmann response $f_0=f_{M}\exp(-Ze\Phi_1/T)$.}

\begin{figure}
\begin{center}
  \includegraphics[width=0.6\textwidth,angle=-90]{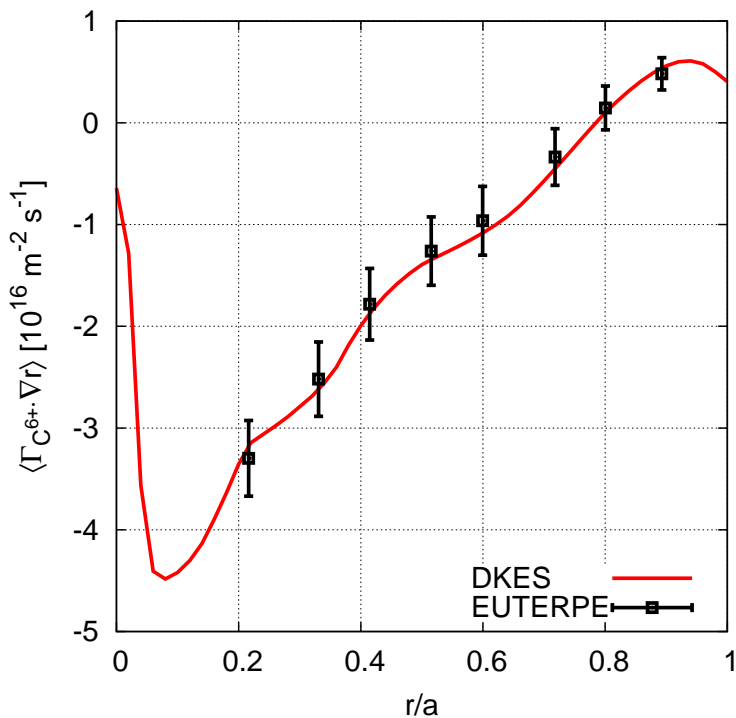}
  \caption{EUTERPE/DKES comparison for the radial particle flux density of C$^{6+}$ as a function of the effective radius
normalized to the minor radius $r/a$. The plasma parameters correspond to the case \text{B.III} shown in fig. \ref{fig:lhd_profiles}.}
  \label{fig:bench_c6}
\end{center}
\end{figure}

The set of eqs. \ref{eq:dotR}-\ref{eq:vlasov} in the 
case $\Phi_{1}$ were neglected would lead
 to that of the local and mono-energetic limit of the neoclassical theory.
In fig. \ref{fig:bench_c6} it is shown a radial profile of 
the particle flux density of C$^{6+}$
is such limit, comparing the result obtained using this version of
EUTERPE with that obtained
with a transport code based on the DKES. 
The case corresponds to the LHD-like magnetic 
configuration considered in the forthcoming section \ref{sec:lhd} for the case labeled as \text{B.III} 
(see figs. \ref{fig:lhd_profiles}),\\

Regarding the calculation of $\Phi_{1}$, instead of the less restrictive ambipolariy
constraint $\sum_{\alpha}Z_{\alpha}e\left<\Gamma_{\alpha}\cdot\nabla r\right>=0$ that determines the ambipolar part of the electrostatic potential $\Phi_{0}$, 
the calculation of $\Phi_{1}$ requires the fulfillment of quasi-neutrality among 
all the species ${\alpha}$,

\begin{equation}
\sum_{\alpha}q_{\alpha}n_{\alpha}=0.
\end{equation}

Taking the density for the species $\alpha$ up to first order as the sum of 
the equilibrium part $n_{\alpha 0}$ (now varying on the flux surface due to the adiabatic response)
and the first order departure

\begin{equation}
n_{\alpha}=n_{\alpha 0}\exp(-Z_{\alpha}e\Phi_{1}/T_{\alpha})+n_{\alpha 1},
\end{equation}

yields to the following relation in the approximation $Ze\Phi_{1}/T \ll 1$

\begin{equation}
\label{eq:qngeneral}
\Phi_{1}=\frac{T_{e}}{e}
\left(n_{0e}+n_{0i}\frac{T_{e}}{T_{i}}+Z^{2}n_{0Z}\frac{T_{e}}{T_{Z}}\right)^{-1}
\left(n_{1i}+n_{1e}+Zn_{1Z}\right).
\end{equation}

For the calculations carried out in this work the electron response has been assumed adiabatic.
In addition the concentration is set sufficiently low in all cases ($Z_{\text{eff}}=1.1$) 
to justify the tracer impurity limit taken into account and neglect the contribution 
of $n_{1Z}$ to $\Phi_1$. The relaxation of this assumption indeed may conflict
with the truncation applied to the Taylor series of the adiabatic part so to keep it 
linear in $\Phi_1$ for the impurity distribution function since, as it is shown in 
next section $Z_{\alpha}e\Phi_{1}/T_{\alpha}$ can become of order unity. 
Considering more terms in the expansion of the Boltzmann response would lead then to a non linear
quasi-neutrality equation, whose complexity if left out of the scope 
of this paper. Thust, under the assumptions aforementioned the resulting equation for $\Phi_1$ reduces to,

\begin{equation}
\label{eq:qngeneral}
\Phi_{1}=\frac{T_{e}}{e}
\left(n_{0e}+n_{0i}\frac{T_{e}}{T_{i}}\right)^{-1}
n_{1i},
\end{equation}

and reflects that the potential required to balance the lack 
of charge quasi-neutrality at zeroth order is guaranteed by the 
first order depature from the equilibrium density of the 
bulk ion species, $H^{+}$ throughout all the work.





\begin{figure}
\begin{center}
  \includegraphics[width=0.3\textwidth]{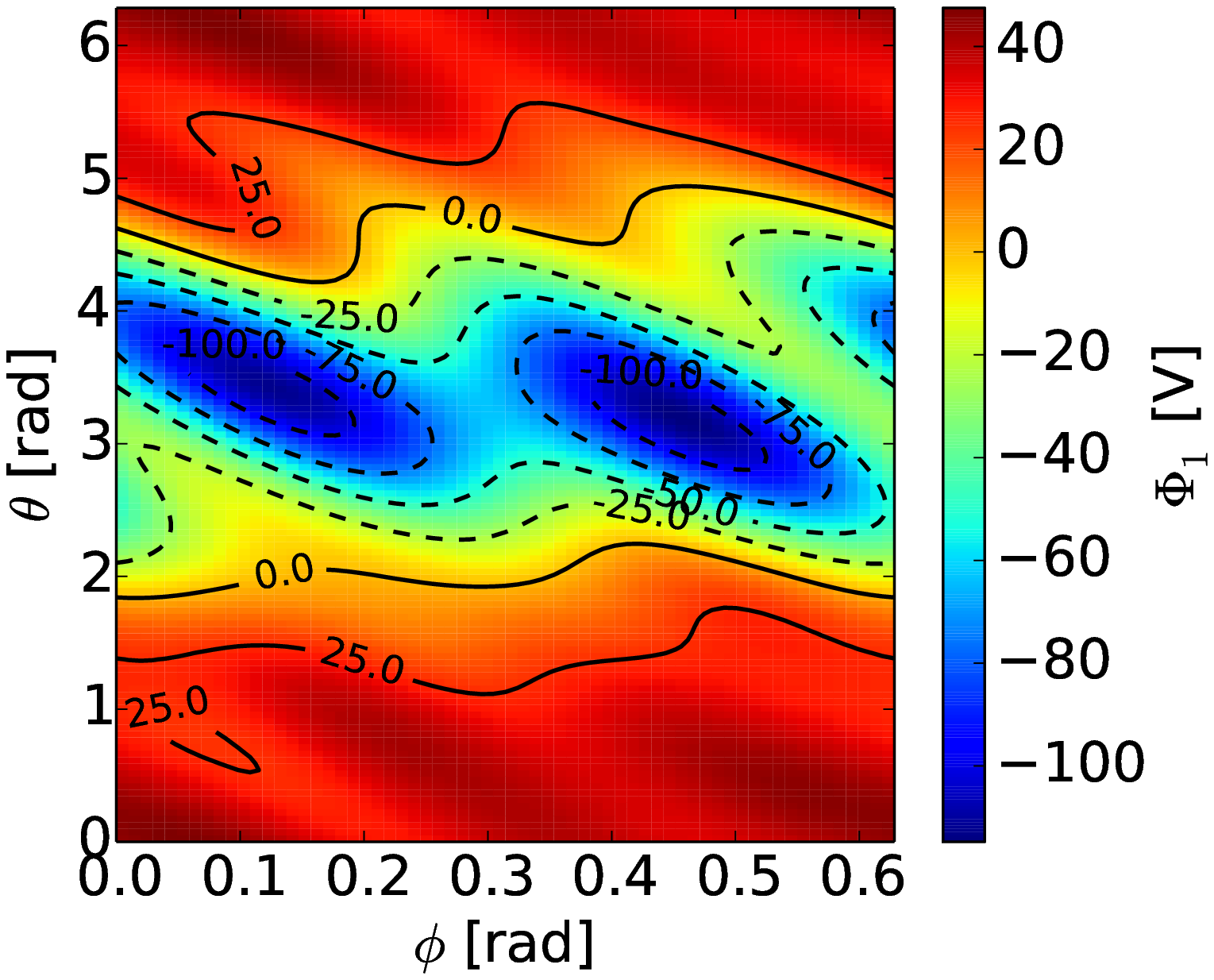}
  \includegraphics[width=0.3\textwidth]{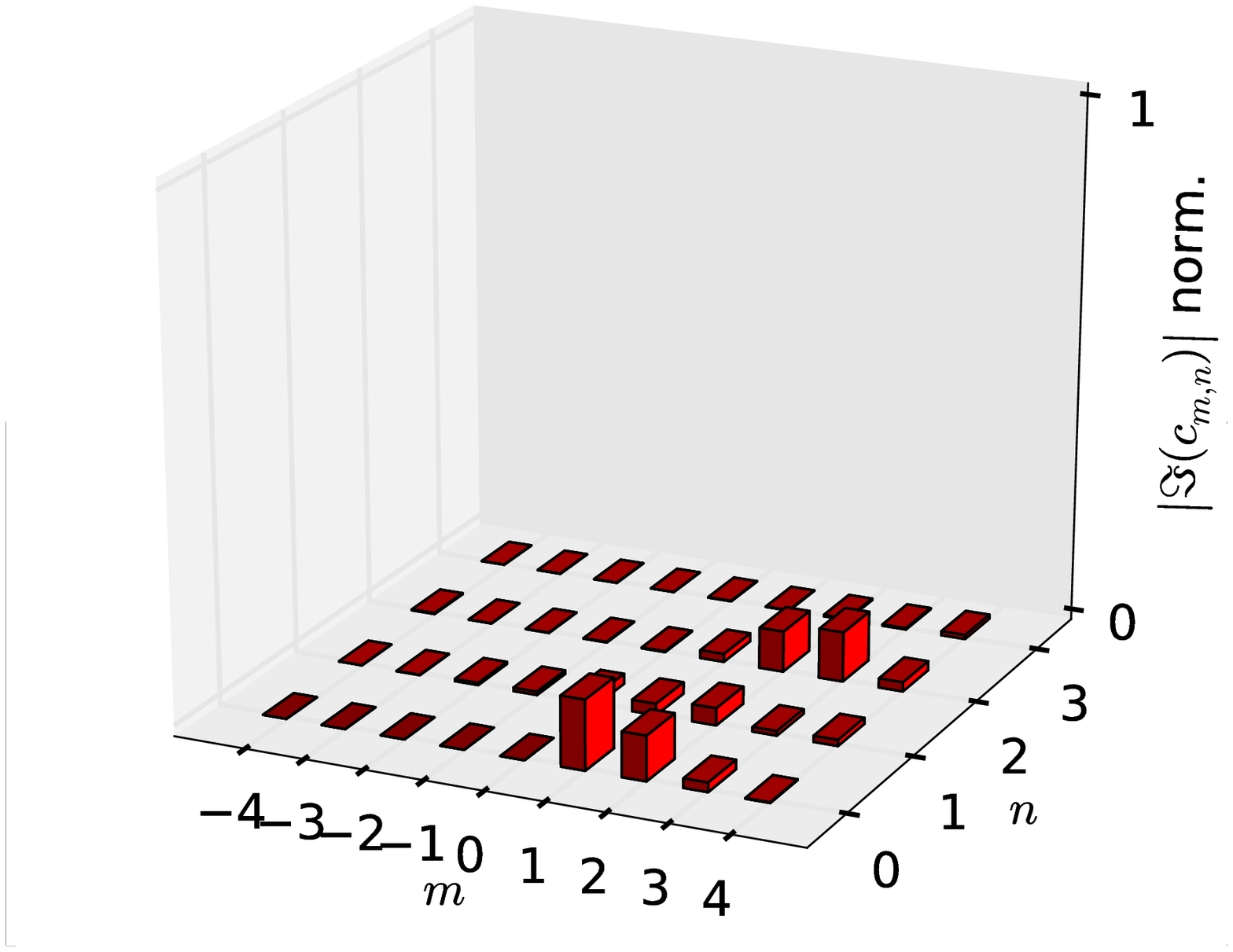}
  \includegraphics[width=0.3\textwidth]{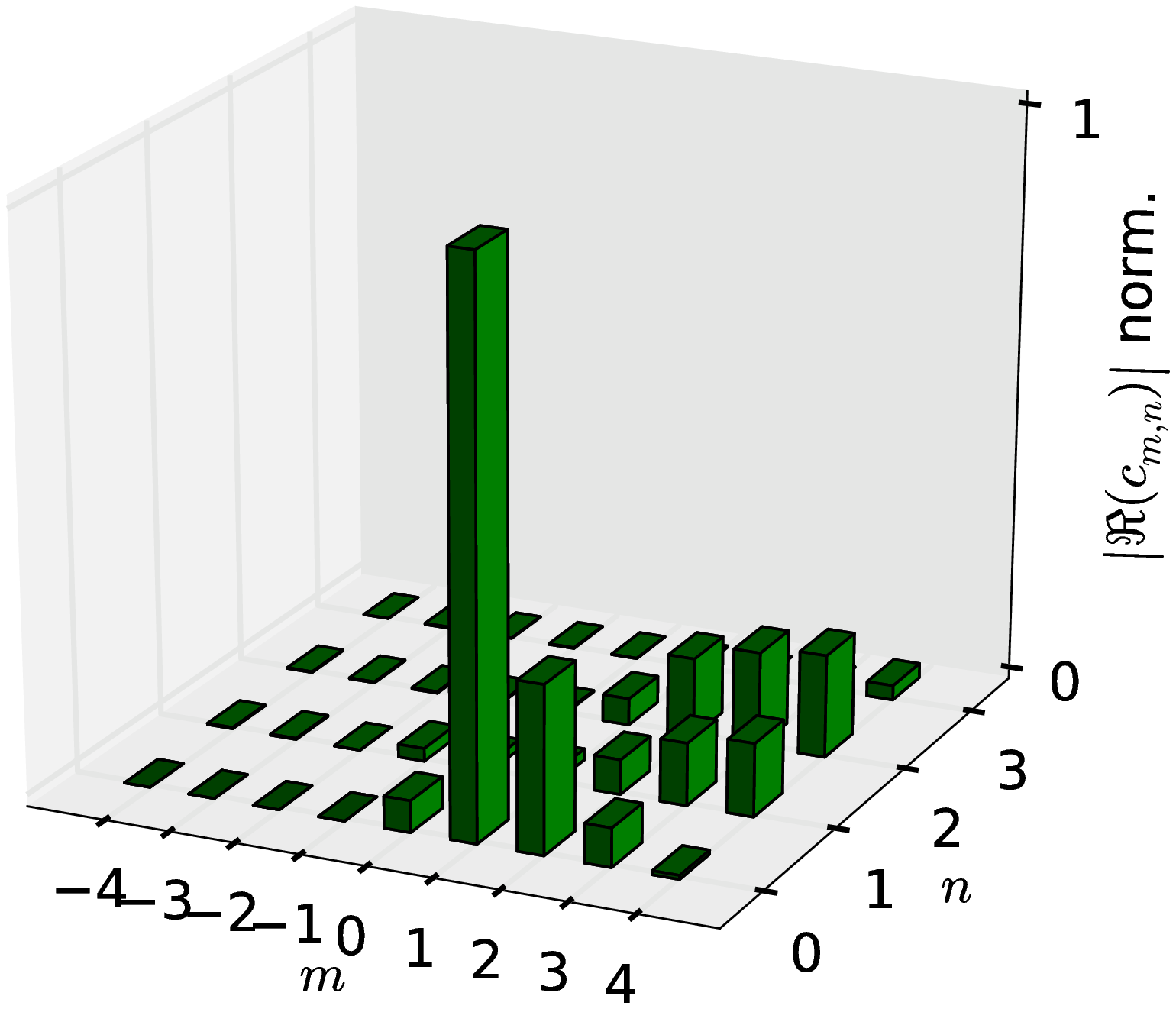}
  \caption{(Left) Stationary solution of $\Phi_{1}$ obtained with EUTERPE 
as a function of the poloidal and toroidal coordinates $\theta$ and $\phi$. 
Real and imaginary parts (left and right respectively) of the Fourier coefficients 
$\Phi_{1,mn}$ on the filtered modes window, and normalized to the 
modulus of the largest component, in this case $\cos\theta$.}
  \label{fig:lhd_example}
\end{center}
\end{figure}

On the numerical side, the Fourier solver implemented in EUTERPE applies 
a low mode number filter to solve the quasi-neutrality equation that
selects the modes in the intervals $[-4,4]$ in 
the poloidal mode number $m$, and $[0,4]$ in the toroidal one $n$.
Note that $\Phi_{1}(\theta,\phi)$ is obtained 
at each flux surface separately. As it is different at each of them
a radial dependency that is not taken into account in the Vlasov equation \ref{eq:vlasov}
is present. 
Furthermore, since $\Phi_{1}$ is obtained at every time step makes that
formally the dependence of $\Phi_{1}$ is more correctly
expressed as $\Phi_{1}(r,\theta,\phi,t)$. Nevertheless, on the one hand the radial 
dependency can be neglected since the radial electric field
related to $\Phi_{1}$ is negligible respect to the ambipolar part:
$\partial\Phi_{1}/\partial t\ll \text{d}\Phi_{0}/\text{d} r$. 
And regarding the time dependency, this is averaged out in the time interval 
when the stationary conditions have been reached. Then, 
for the  remeinder of this paper we will assume the dependence of $\Phi_{1}$ 
as it is actually felt by the tracer impurities in our calculations of their fluxes, 
just a stationary potential $\Phi_{1}(\theta,\phi)$ plugged into eq. \ref{eq:vlasov}
to which impurities do not contribute given its low concentration. Thus
the nonlinear feedback of the impurity 
density inhomogeneity on themselves through its impact on $\Phi_{1}$ is kept out of the scope of this work as well. 
An example of such map with the corresponding spectrum represented with the absolute values 
of the real and imaginary parts of the complex Fourier coefficients normalized to the 
component with maximum amplitude is shown in fig. \ref{fig:lhd_example}. We advance that these example
figure correspond to the case \textit{B.III} discussed later in section \ref{sec:lhd}.

\section{Radial flux of C$^{6+}$ including $\Phi_{1}$}
\label{sec:results}

In this section the numerical results are presented and discussed.
Three different non-axisymmetric devices have been considered: the heliotron type Large Helical Device (LHD, Toki, Japan); 
the helias type Wendelstein 7-X (W7-X, Greifswald, Germany) and the heliac TJ-II (Madrid, Spain). 
For each of them a vacuum configuration has been used and the radial
particle flux of C$^{6+}$ calculated, comparing the result 
when $\Phi_1$ is neglected with the result when $\Phi_1$ is taken into account.
The plasma parameters scanned have resulted in 6 different sets of 
density and temperature profiles for LHD, 4 for W7-X and 2 for TJ-II.
The paramenters related to the magnetic configuration for these 
three devices are listed in table \ref{tab:parameters}.

\begin{table}
\begin{center}
\begin{tabular}{ c c c c }
  \hline
  \multicolumn{4}{c}{\textbf{Magnetic configurations}}\\
  \hline
  Device & $B_{0,0} (r/a=0.5)$ [T]& $R$ [m] & $a$ [a] \\
  \hline
  LHD & 1.54 & 3.6577 & 0.5909 \\
  Wendelstein 7-X & 2.78 & 5.5118 & 0.5129 \\
  TJ-II & 0.996 & 1.5041 & 0.1926\\
  \hline
\end{tabular}
\caption{Major and minor radii, $R$ and $a$ respectively, and amplitude of the  Boozer harmonic $(m,n)=(0,0)$ of the magnetic field at the mid plasma radius, $B_{0,0}$.}\label{tab:parameters}
\end{center}
\end{table}

\subsection{LHD results}
\label{sec:lhd}

\begin{figure}
\begin{center}
  \includegraphics[width=0.3\textwidth]{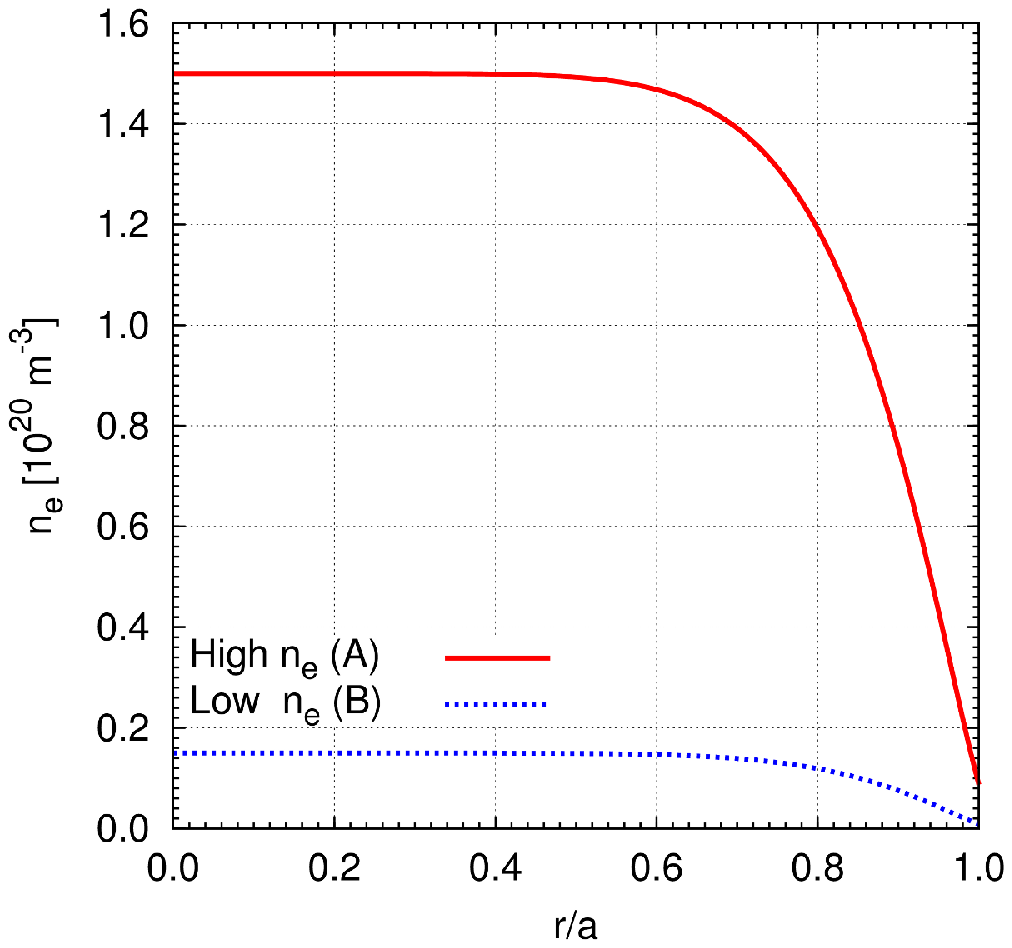}
  \includegraphics[width=0.3\textwidth]{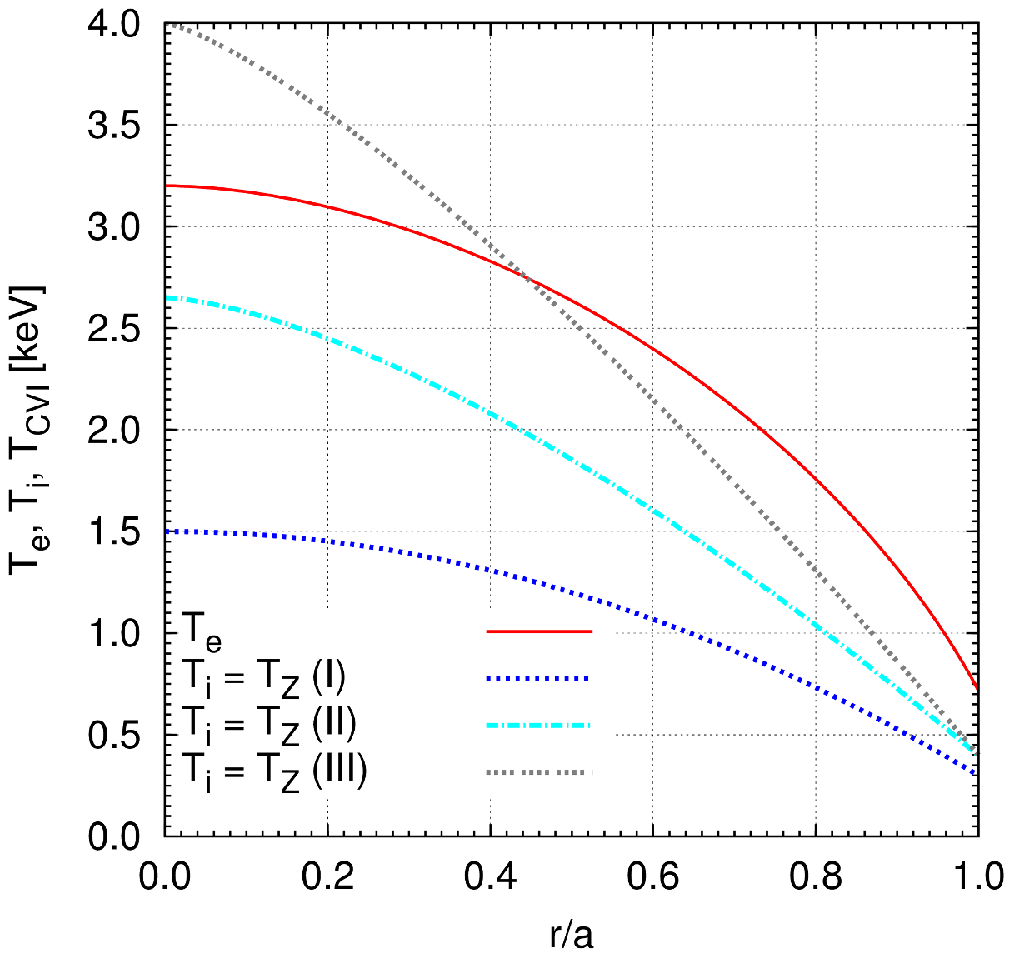}
  \includegraphics[width=0.33\textwidth]{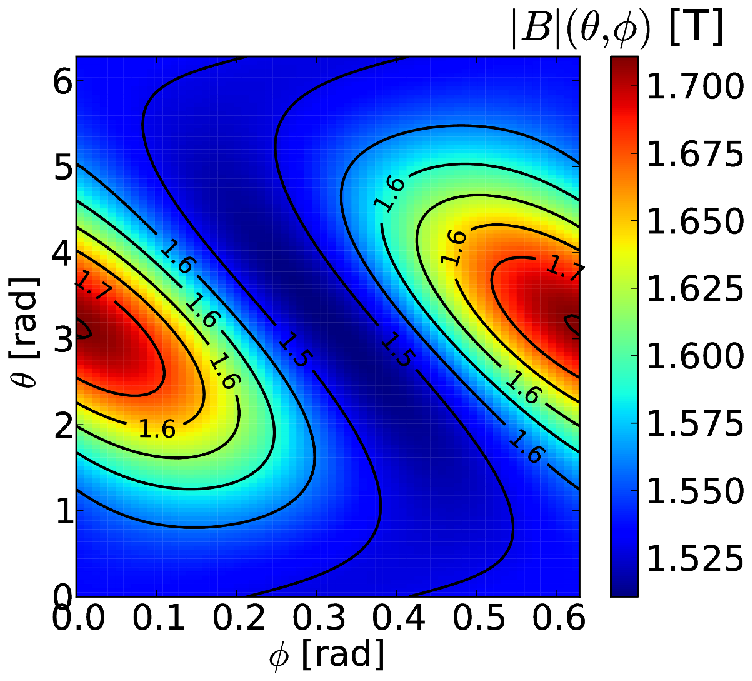}  
  \caption{Set of profiles considered for the LHD cases. (Left) Electron density profiles. 
(Center) Electron, Hydrogen and C$^{6+}$ temperature profiles. (Right) Magnetic field 
modulus in one LHD field period at the radial position $r/a=0.5$.}
  \label{fig:lhd_profiles}
\end{center}
\end{figure}

For the LHD two different density profiles have been considered. A first one 
corresponding to a high density scenario and a second one corresponding to a low 
density one. 
These two profiles are represented in fig. \ref{fig:lhd_profiles} (left)
and labeled respectively as \textit{A} (high density) and \textit{B} (low
density). \blue{The equilibrium density $n_{0i}$
of the bulk ions (Hydrogen nuclei) and impurities $n_{0Z}$ -- in all cases fully ionized 
Carbon C$^{6+}$ -- is determined so to fulfill quasi-neutrality among these 
and the electrons: $\sum_{\alpha}Z_{\alpha}e n_{0\alpha}=0$. The reference value
considered for the effective charge is for all cases 
$Z_{\text{eff}}=1.1$, that allows to assume negligible the impact of 
the radial flux of Carbon on the ambipolar electrostatic potential $\Phi_0$,
and of the influence of the impurity perturbed density $n_{1Z}$ on the potential $\Phi_1$.}
Three temperature profiles for C$^{6+}$, set equal to the bulk ion temperature $T_{i}$,
have been used, letting the electron
temperature $T_{e}$ profiles fixed. These profiles, sorted by increasing ion temperature
are labeled as \textit{I}, \textit{II}
and \textit{III}, and are shown in fig. \ref{fig:lhd_profiles} (center).
Finally the magnetic field strength of the magnetic 
configuration is at the position $r/a = 0.6$ illustrated in fig. \ref{fig:lhd_profiles} (right)
as a function of the poloidal and toroidal angular coordinates $\theta$ and $\phi$.\\

\begin{figure}
\begin{center}
  \includegraphics[width=0.3\textwidth]{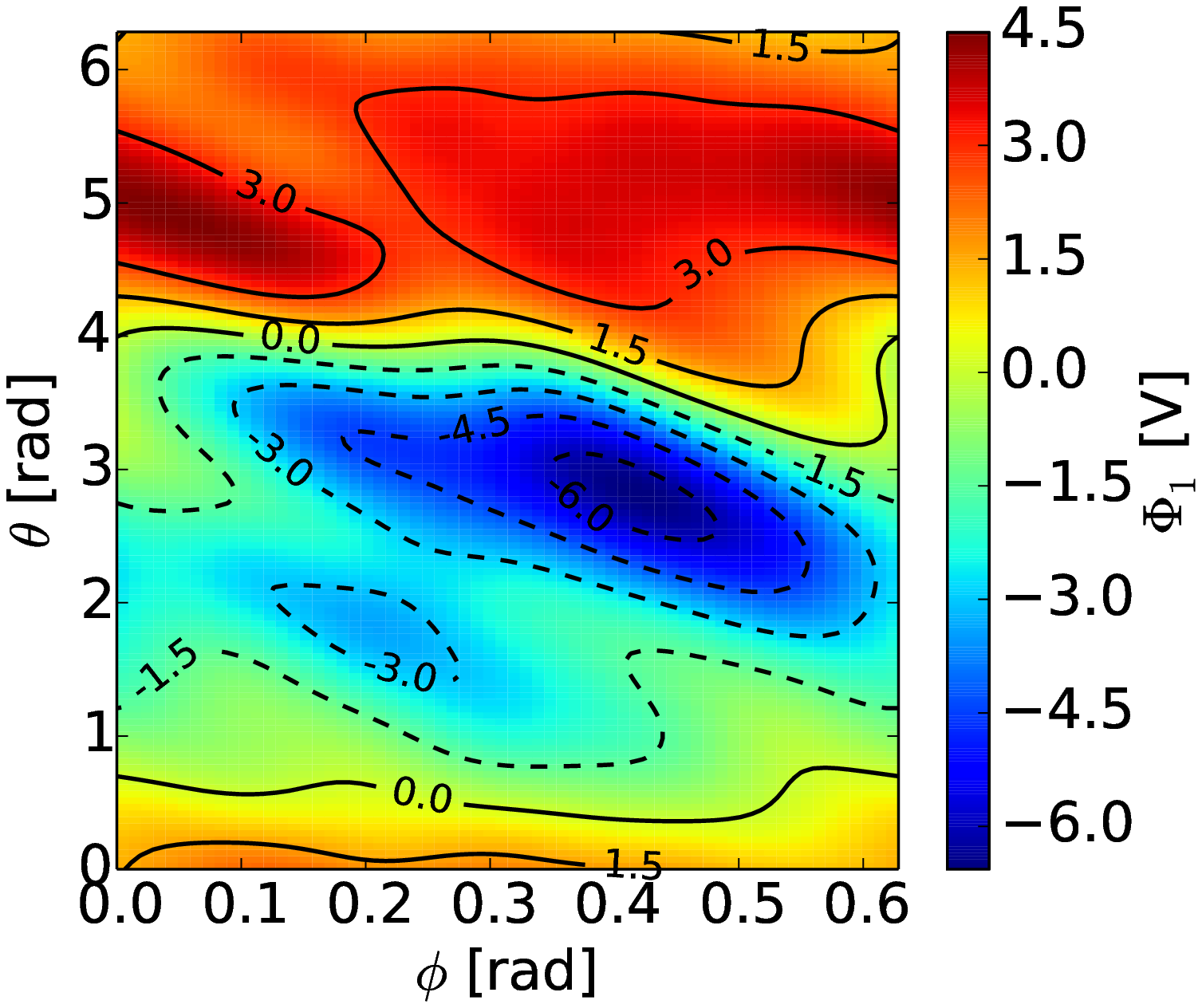}
  \includegraphics[width=0.3\textwidth]{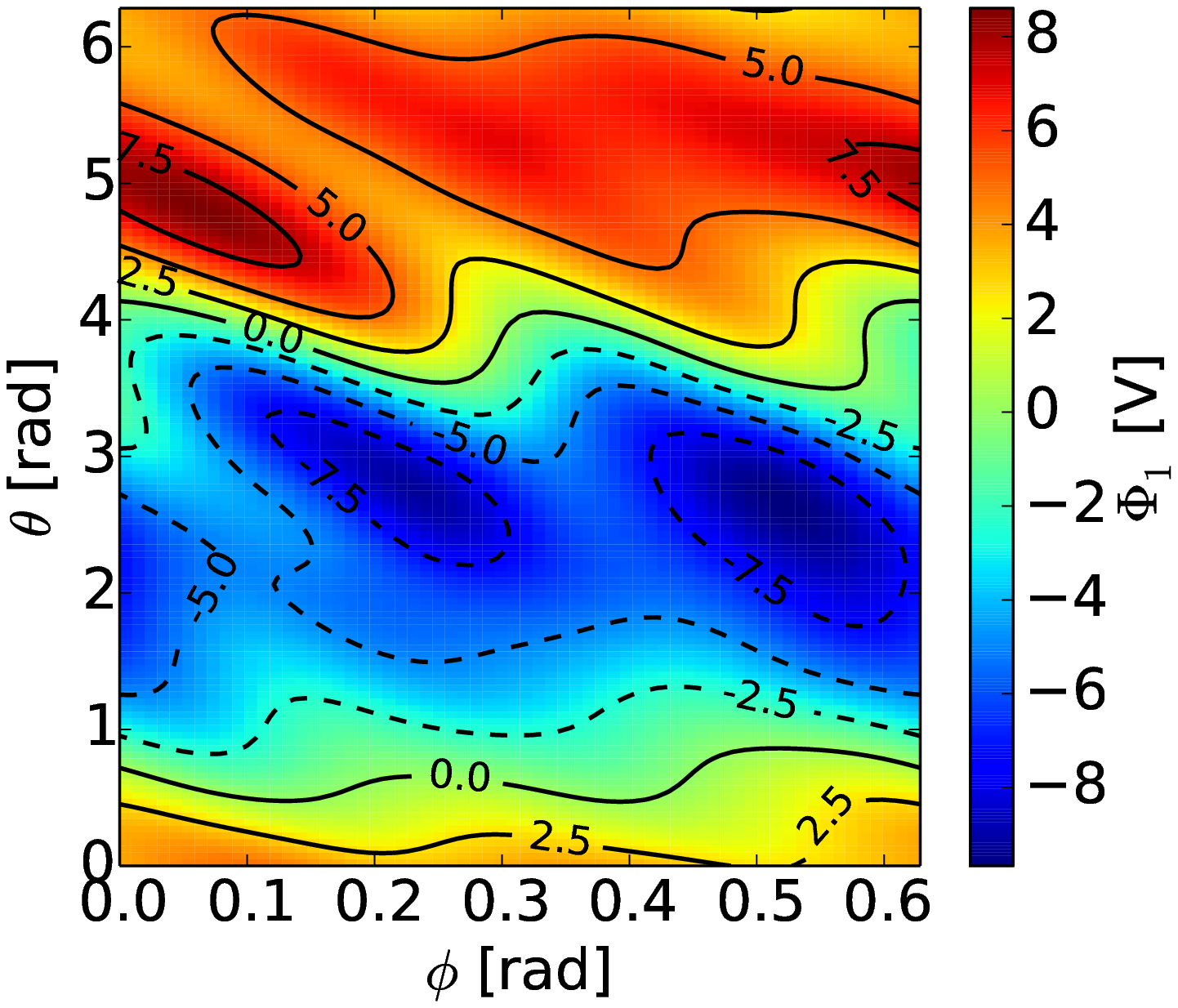}
  \includegraphics[width=0.3\textwidth]{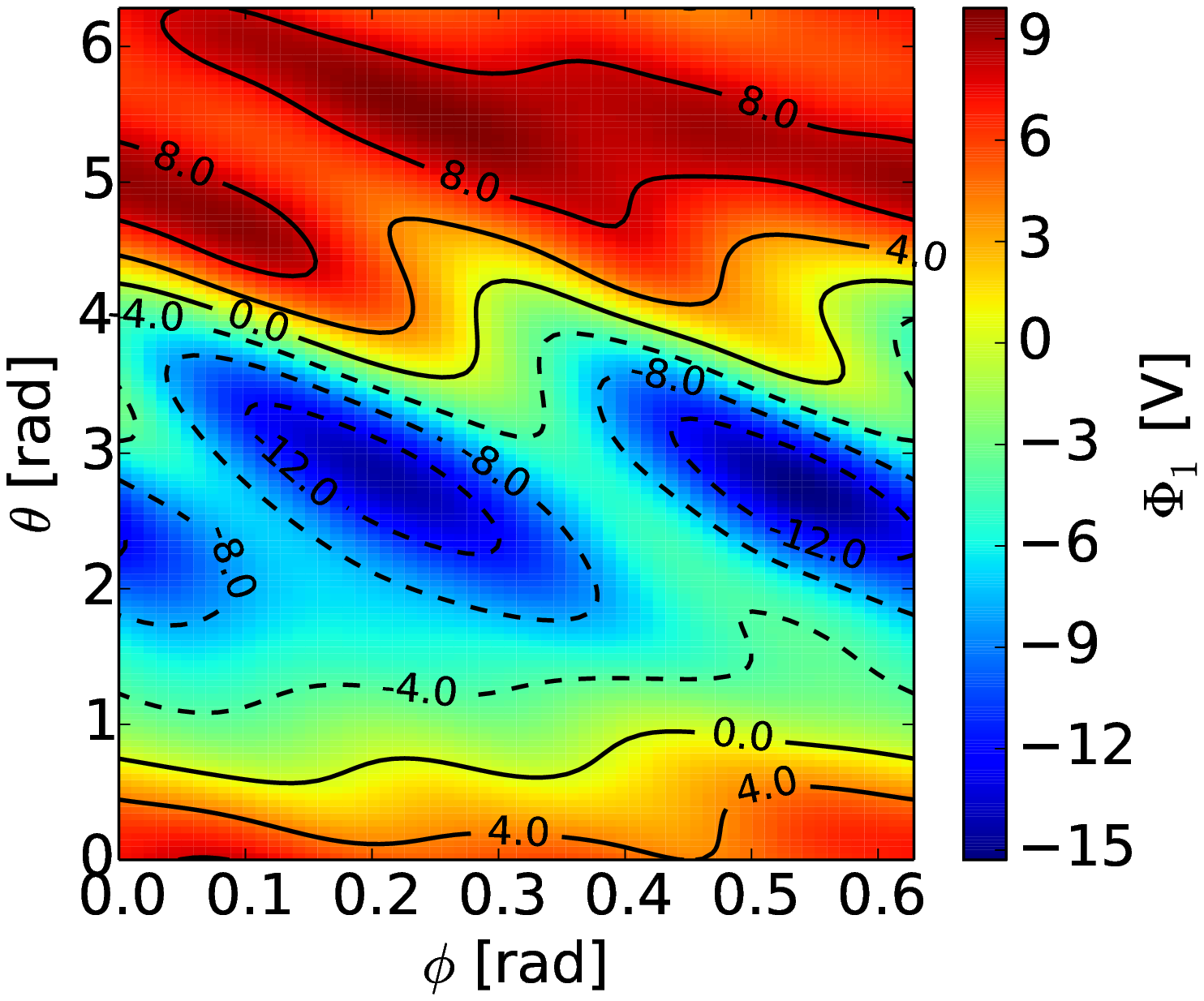}\\
  \includegraphics[width=0.3\textwidth]{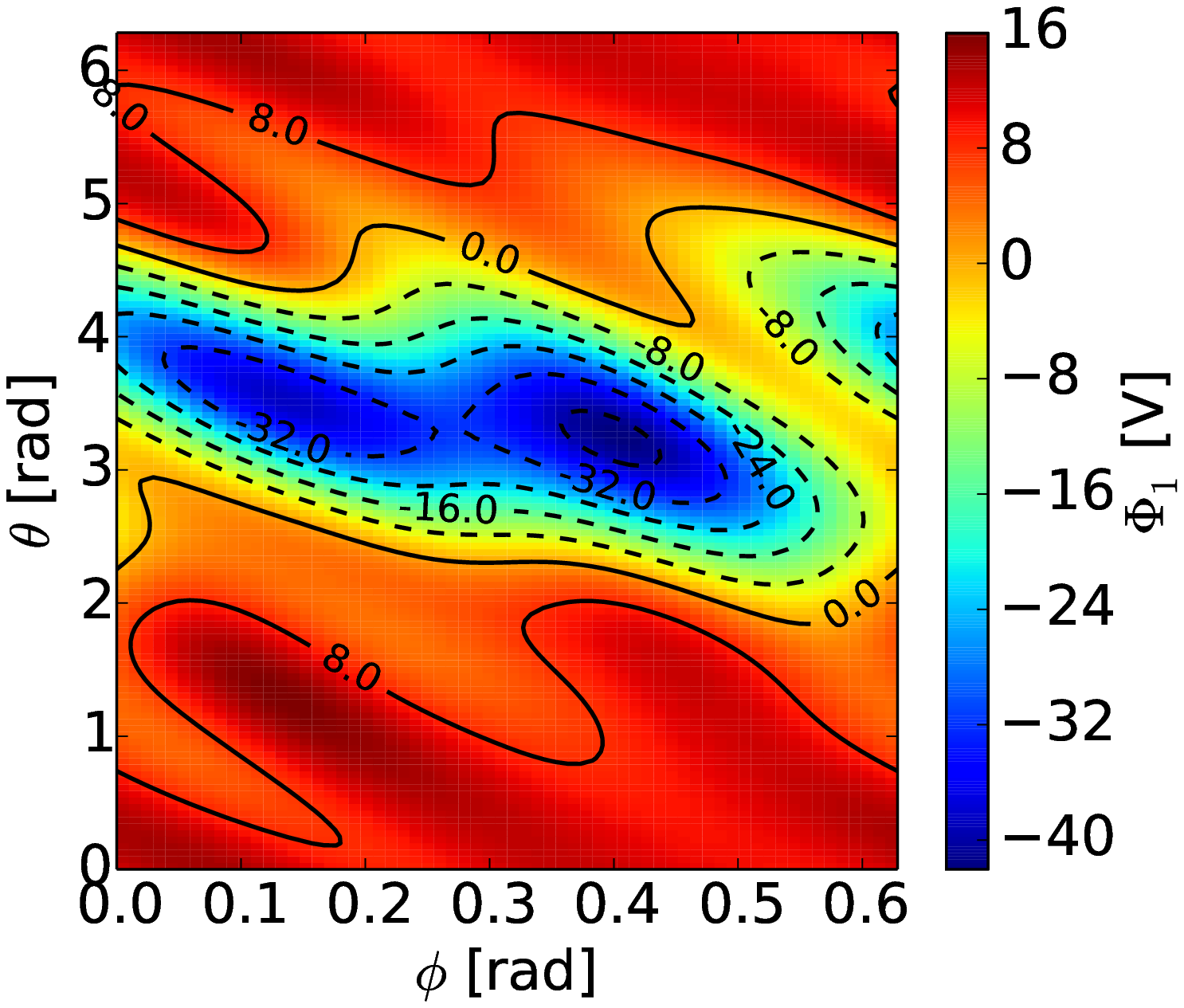}
  \includegraphics[width=0.3\textwidth]{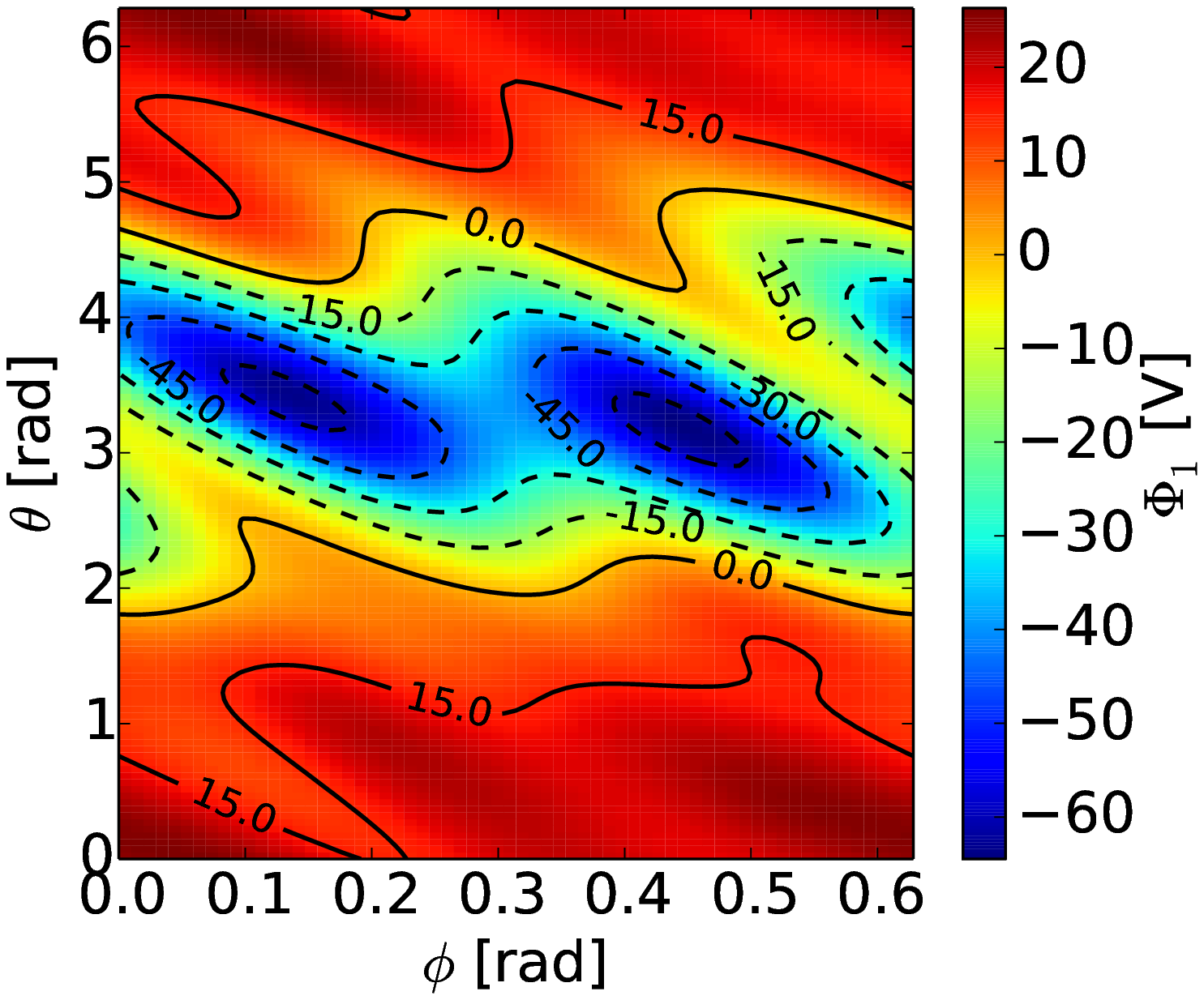}
  \includegraphics[width=0.3\textwidth]{phi2d_helios_lhd2_663.eps}  
  \caption{At the position $r/a=0.6$ for LHD: on the top row $\Phi_{1}(\theta,\phi)$ 
for the high density case (A) and on the bottom row for the low density case (B). 
The left, center and right column corresponds to the temperature profiles considered
I, II and III respectively.}
  \label{fig:lhd_phi1maps}
\end{center}
\end{figure}

In fig. \ref{fig:lhd_phi1maps} for the low density set of profiles \textit{A} 
the electrostatic potential depending on the poloidal and toroidal coordinates 
$\Phi_{1}(\theta,\phi)$ is represented on the top row for the three mentioned 
ion and C$^{6+}$ temperature profiles 
\textit{I}, \textit{II} and \textit{III}. 
Equivalently the bottom row contains for the low density set \textit{B} the
three maps of $\Phi_{1}$ for the same three temperature profiles.
All the six maps are considered at the radial position $r/a=0.6$.\\
It can be observed the qualitative trend mentioned in the 
introduction. As the maps are viewed from left to right 
and top to bottom, this is with monotonically decreasing ion collisionality,
the variation of the potential from peak to peak is increased in one order of magnitude, 
accordingly with the decrease of the collisionality (of a factor 40 approximately). 
This justifies to some extent the neglect of the 
electron contribution to $\Phi_{1}$ we are considering, since in the cases 
considered the electron to ion collisionality is also of the same order larger
than that of the bulk ions. 

\begin{figure}[t]
\begin{center}
  \includegraphics[width=0.3\textwidth]{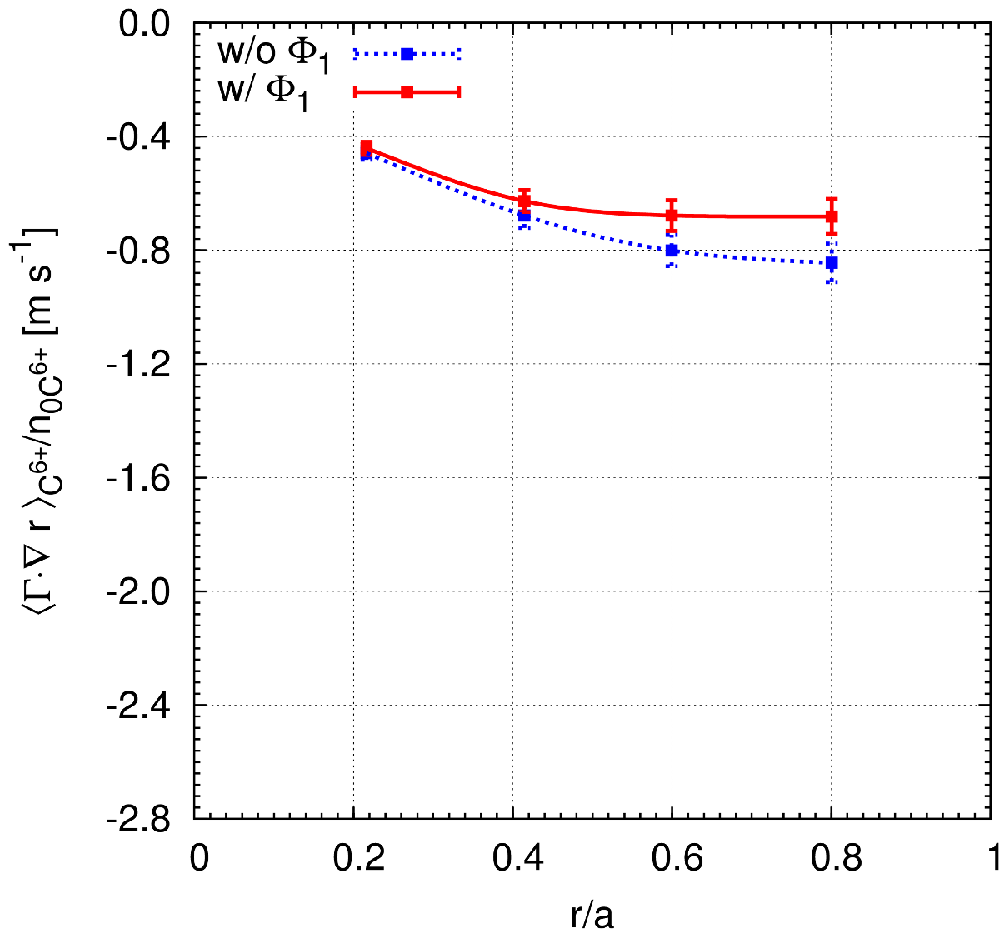}
  \includegraphics[width=0.3\textwidth]{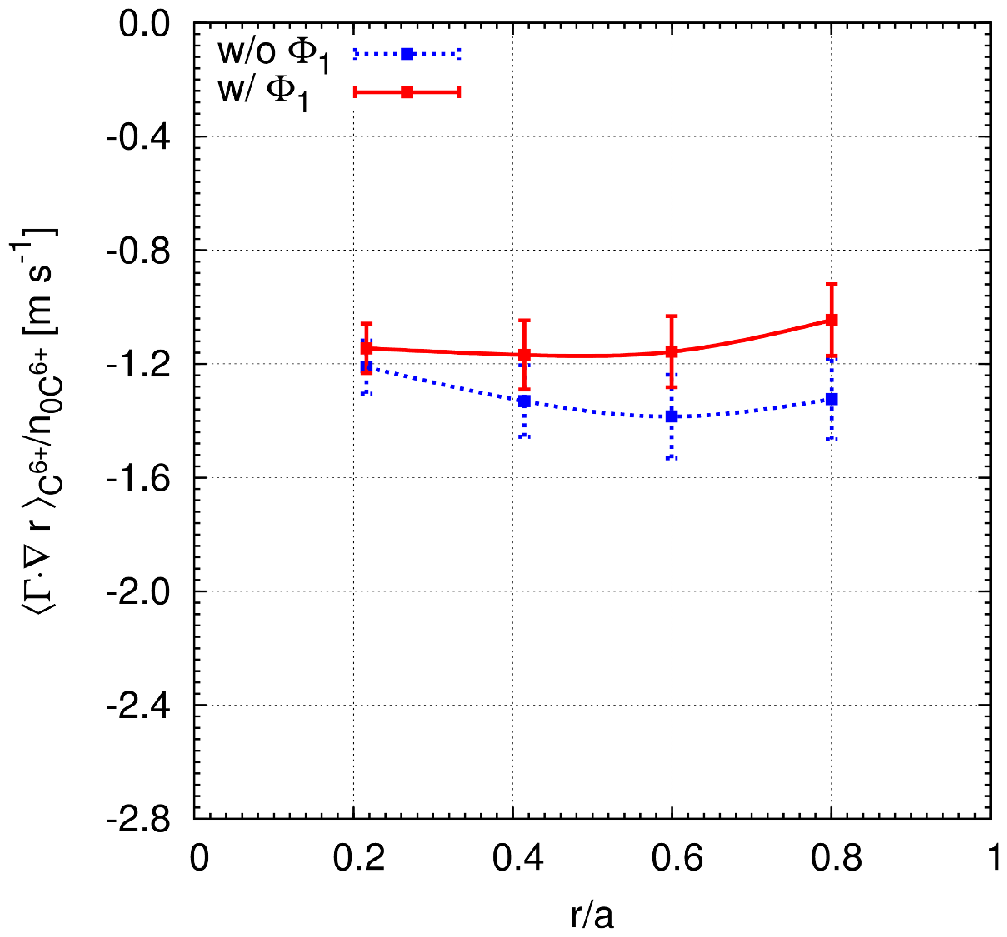}
  \includegraphics[width=0.3\textwidth]{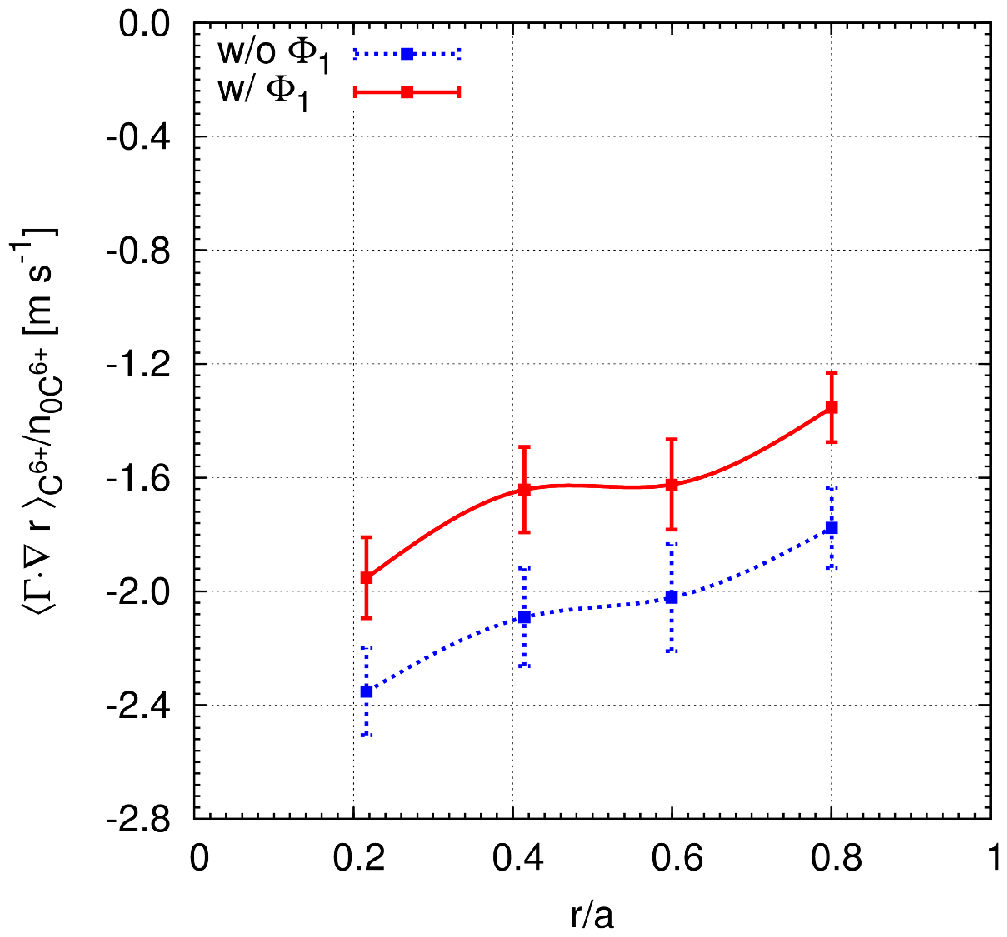}\\
  \includegraphics[width=0.3\textwidth]{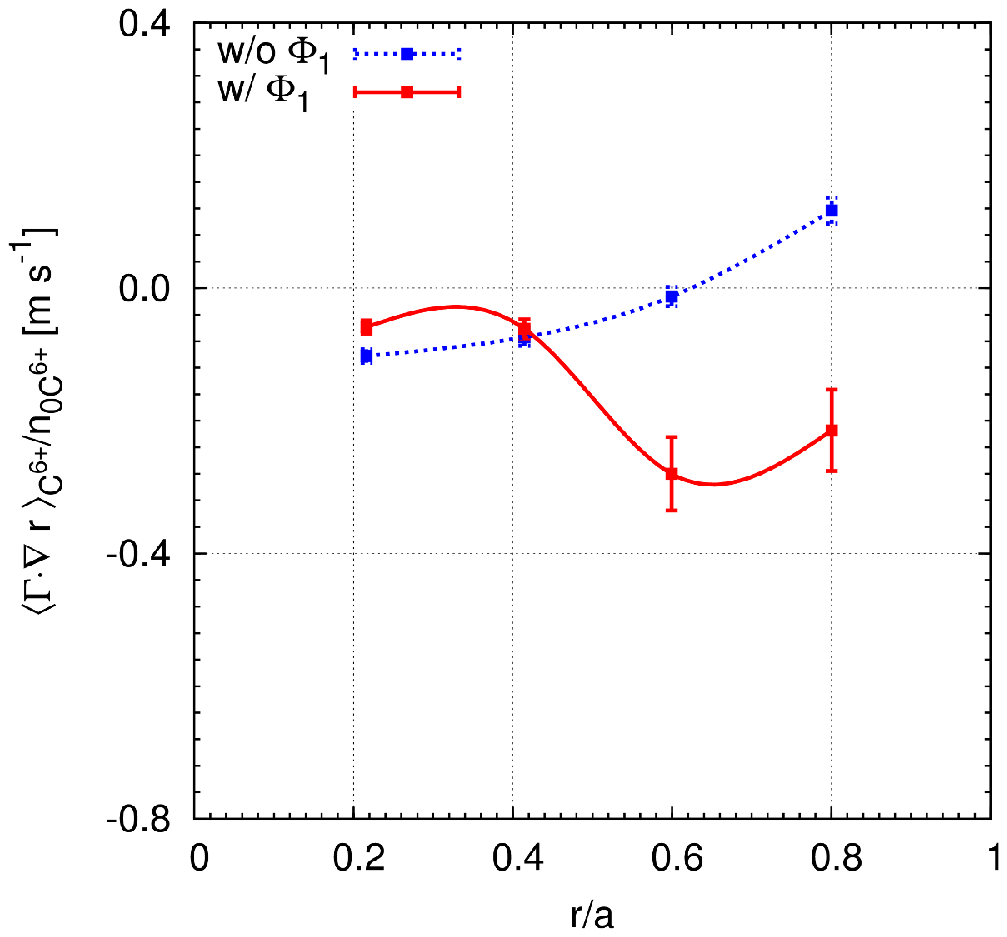}
  \includegraphics[width=0.3\textwidth]{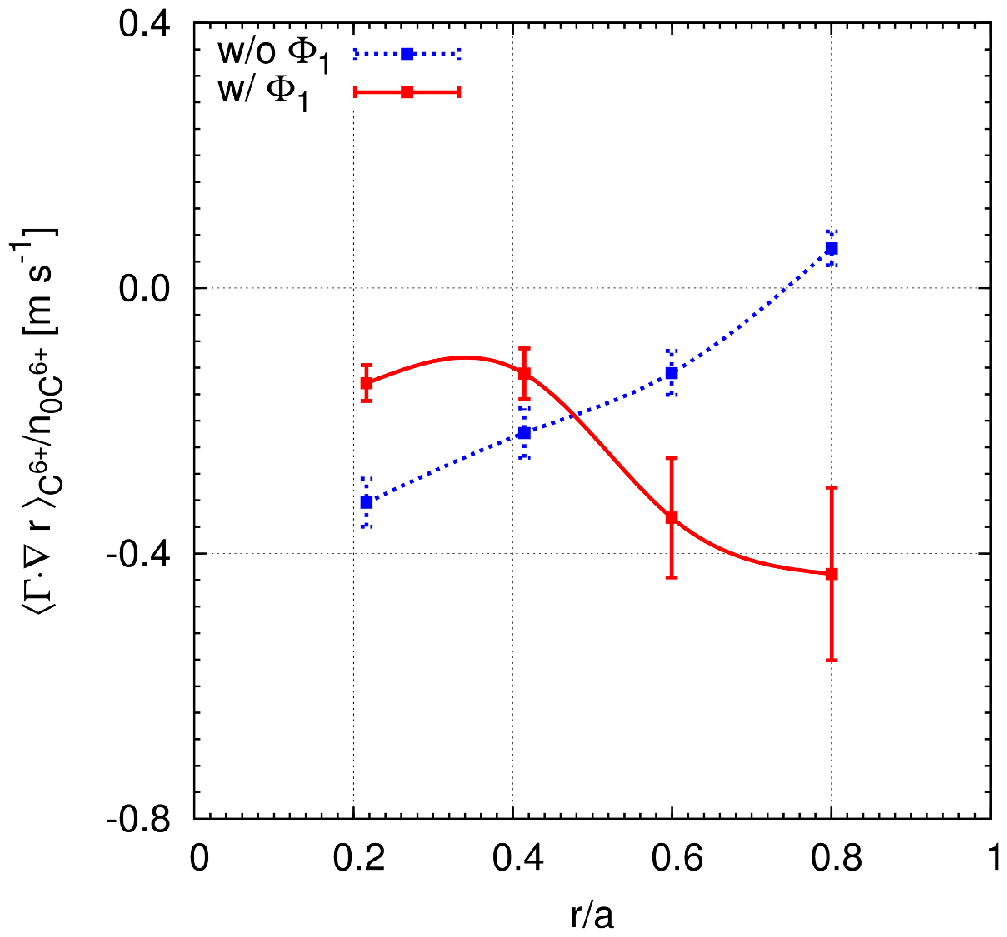}
  \includegraphics[width=0.3\textwidth]{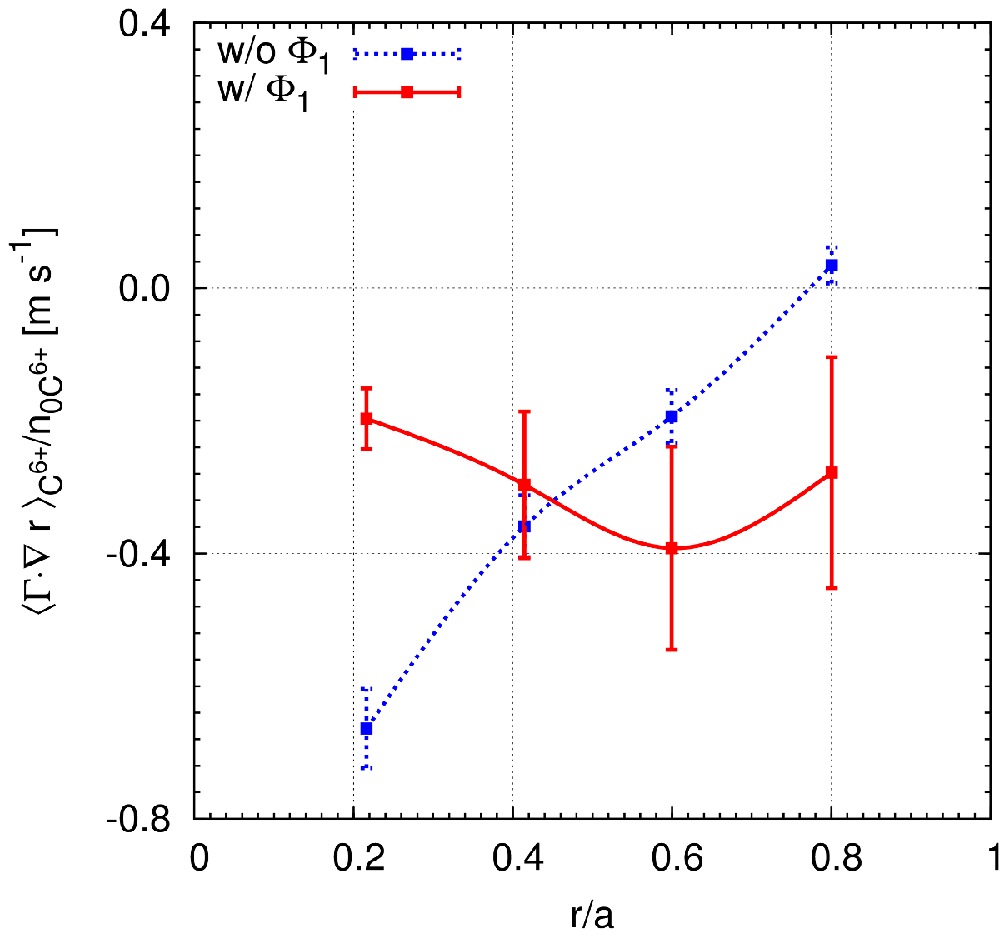}
  \caption{Radial flux density of C$^{6+}$ as a function of $r/a$ including $\Phi_{1}$ (solid line) and 
neglecting it (dotted line). From left to right the temperature profiles I, II and III are considered
and from top to bottom the high \textit{A} and low \textit{B} density cases.}
  \label{fig:lhd_pflux}
\end{center}
\end{figure}

Regarding the central question of how strongly is the impurity transport
affected by $\Phi_{1}$, the potential was calculated at 
the radial positions $r/a=\left\{0.2, 0.4, 0.6, 0.8\right\}$ for each set of profiles, 
and the resulting maps of $\Phi_{1}$ considered as an input for the calculation
of the C$^{6+}$ radial particle flux $\left<\boldsymbol{\Gamma}\cdot\nabla r\right>$.
The radial particle flux density of C$^{6+}$ as a function of the flux surface
label is shown in fig. \ref{fig:lhd_pflux} for each 
of the 6 profiles sets considered in this section, and normalized to the equilibrium 
impurity density $n_{0Z}$. 
The dotted line represents the case where only the ambipolar electric field is taken into account, and the 
solid onee corresponds to the case where both the ambipolar part of the potential $\Phi_{0}$ and 
$\Phi_{1}$ are both taken into account.
Each of the 6 plots is accordingly displayed at the same position
than the maps in fig. \ref{fig:lhd_phi1maps} do refered to the 
pair of profiles considered. This is, on the top row the high density scan \textit{A}
and on the bottom one the low density
one \textit{B} are shown; and from left to right the figures corresponds to the $T_{i}=T_{\text{\tiny{C$^{6+}$}}}$ profile \textit{I} (left), 
\textit{II} (center) and \textit{III} (right).\\

\blue{At a first glance, \textit{reading} the plots of fig. \ref{fig:lhd_pflux} 
from left to right and top to bottom and looking at the maps in \ref{fig:lhd_phi1maps}
as orientative for the order of magnitude of $\Phi_1$ for each case 
(note that the maps shown are for the position $r/a=0.6$ only),
it is found that the impact that $\Phi_1$ produces on the ratial particle flux of Carbon 
is continuously increasing as $\Phi_1$ does. And interestingly, the modification 
that this represents respect to the standard neoclassical picture 
does not point always to the same direction. $\Phi_1$ both mitigates and amplifies the trend
of C$^{6+}$ to accumulate. On the one hand, 
in the high density scan (three figures on top) the impact
of $\Phi_1$ results in a weak mitigation of the inwards flux at all radial positions considered, 
almost negligible in the lowest $T_{i}$ case (top, left) but appreciable in the highest $T_{i}$ one (top, right).
On the other hand, the three figures at the bottom corresponding to the low density temperature 
scan coincide to the fact that $\Phi_1$ reduces the inwards Carbon flux
at the internal radial positions below $r/a=0.4$ approximately and makes it stronger from there
outwards, resulting in the less collisional of all the cases considered \textit{B.III} (bottom, right)
in the reversal of the behavior predicted by the standard neoclassical approach and a substantial 
reduction of the inwards flux.}


\begin{figure}[t]
\begin{center}
  \includegraphics[width=0.3\textwidth]{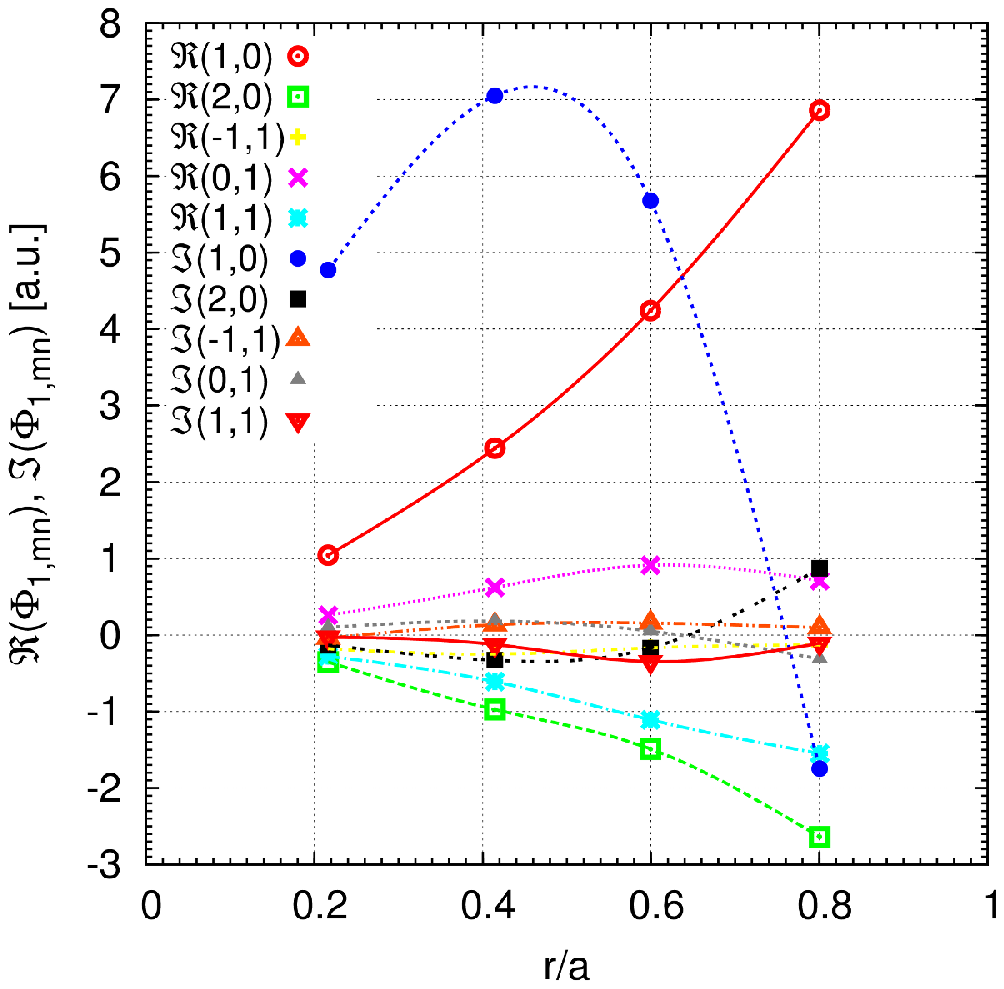}
  \includegraphics[width=0.3\textwidth]{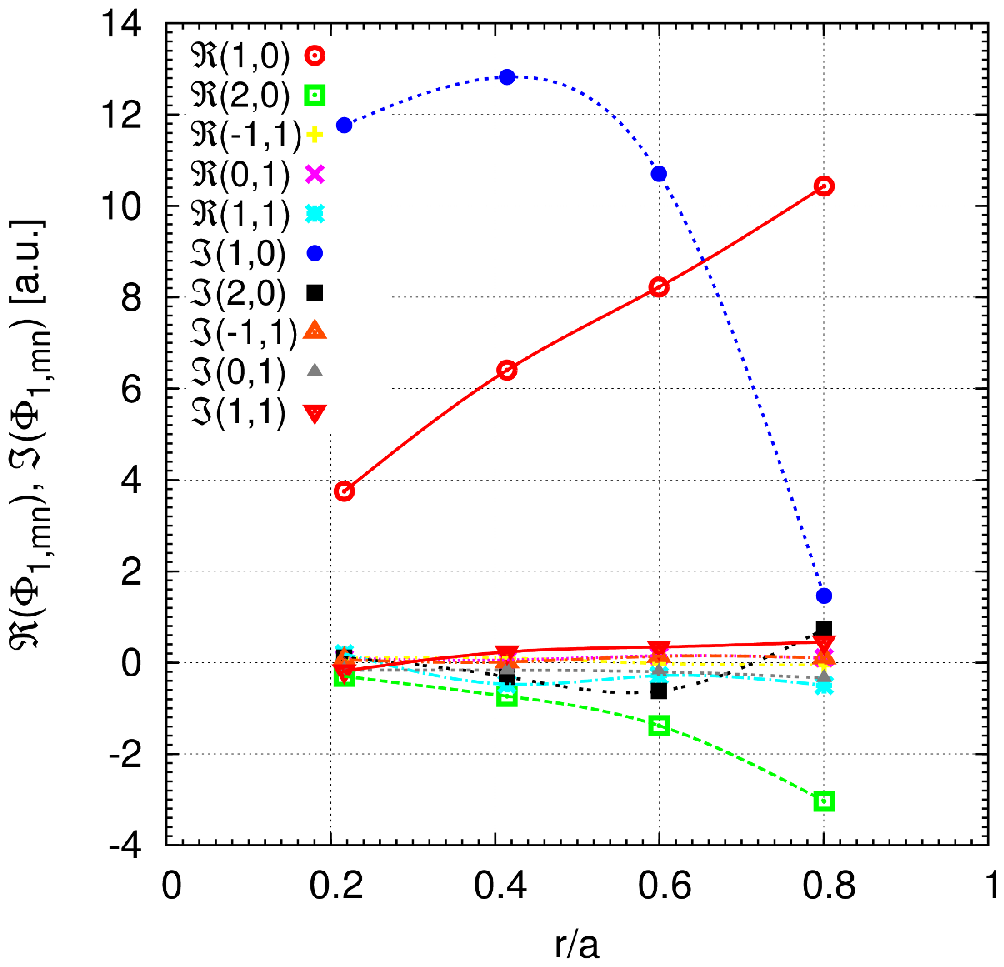}
  \includegraphics[width=0.3\textwidth]{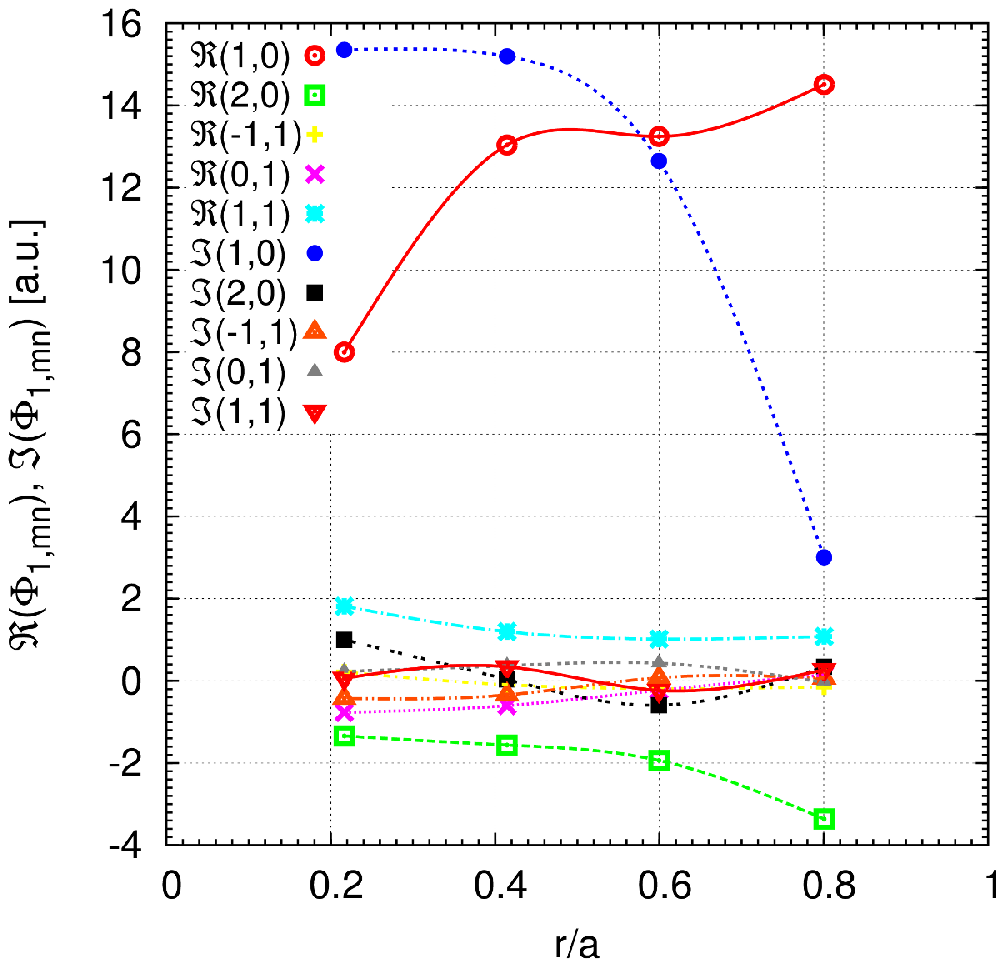}\\
  \includegraphics[width=0.3\textwidth]{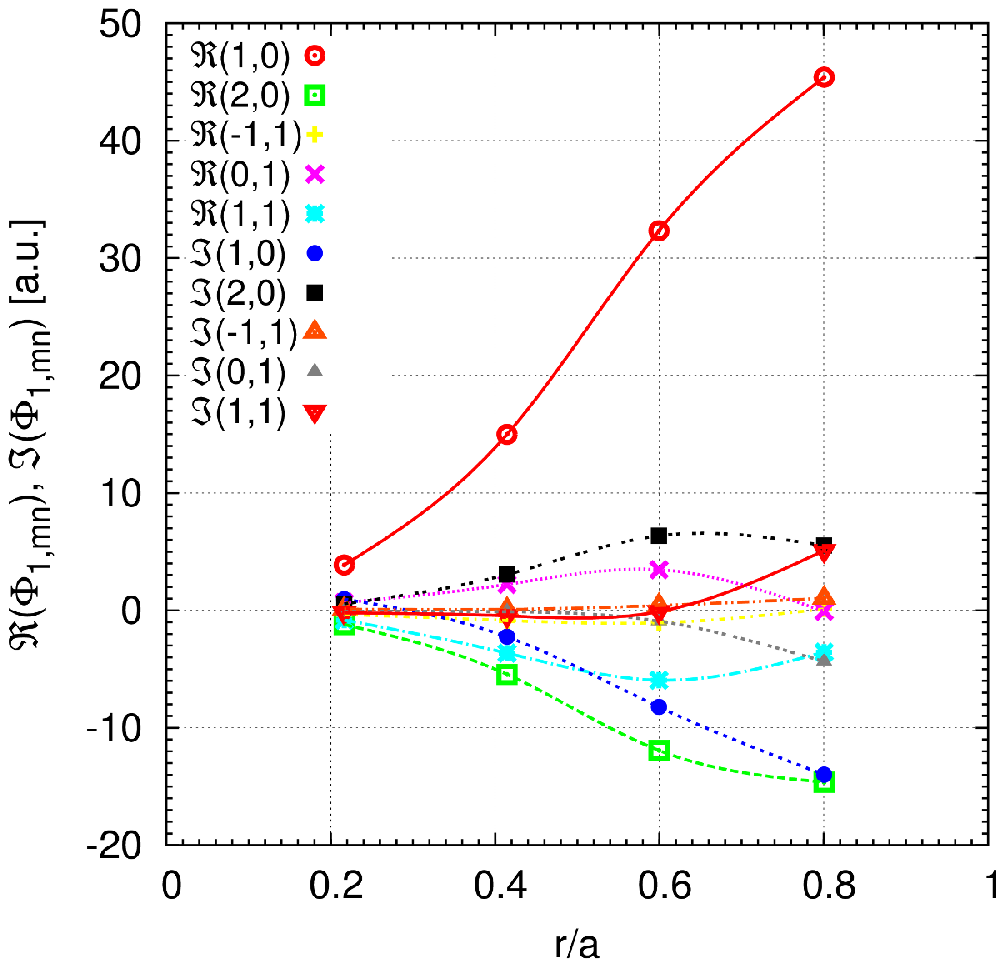}
  \includegraphics[width=0.3\textwidth]{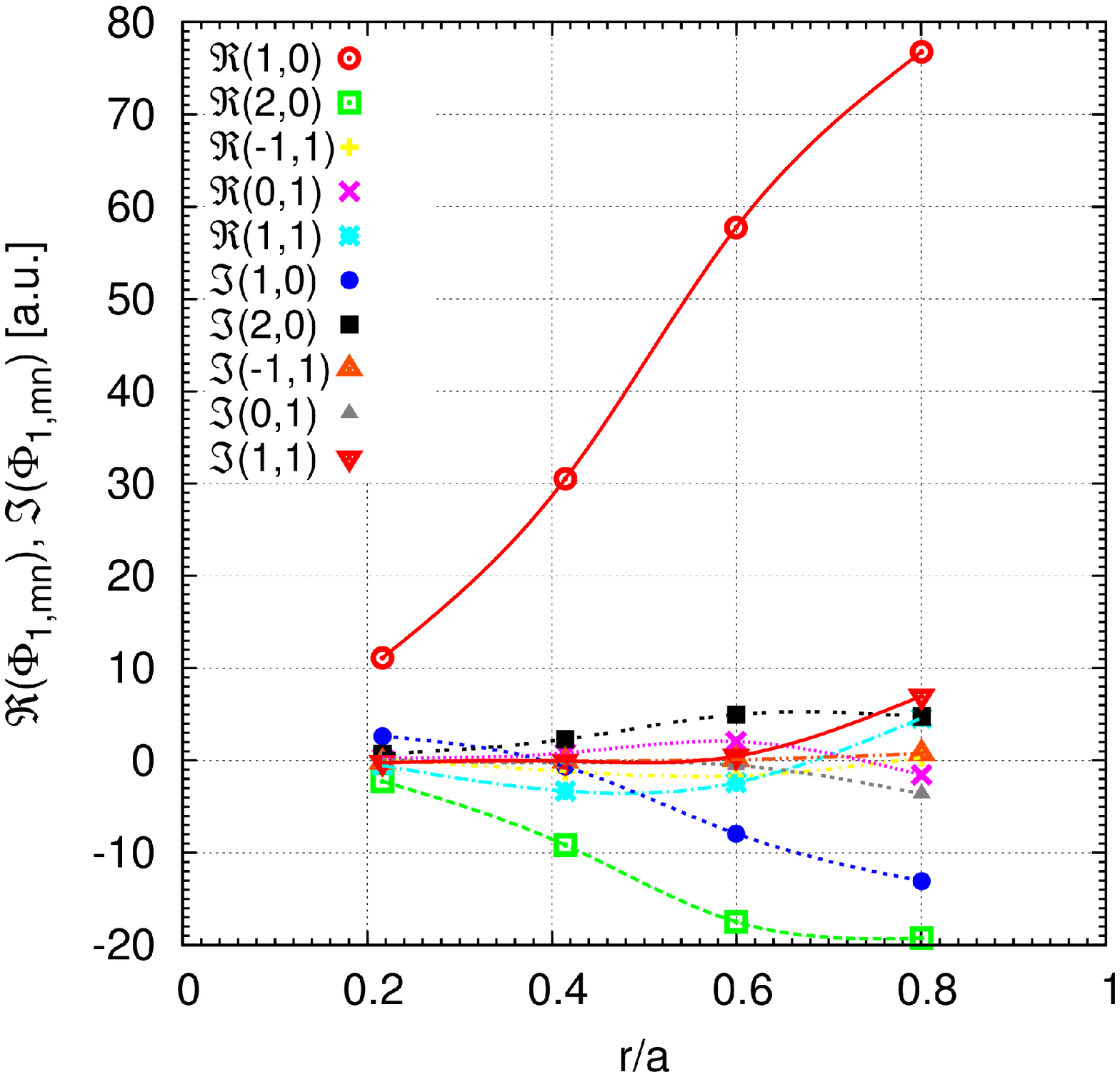}
  \includegraphics[width=0.3\textwidth]{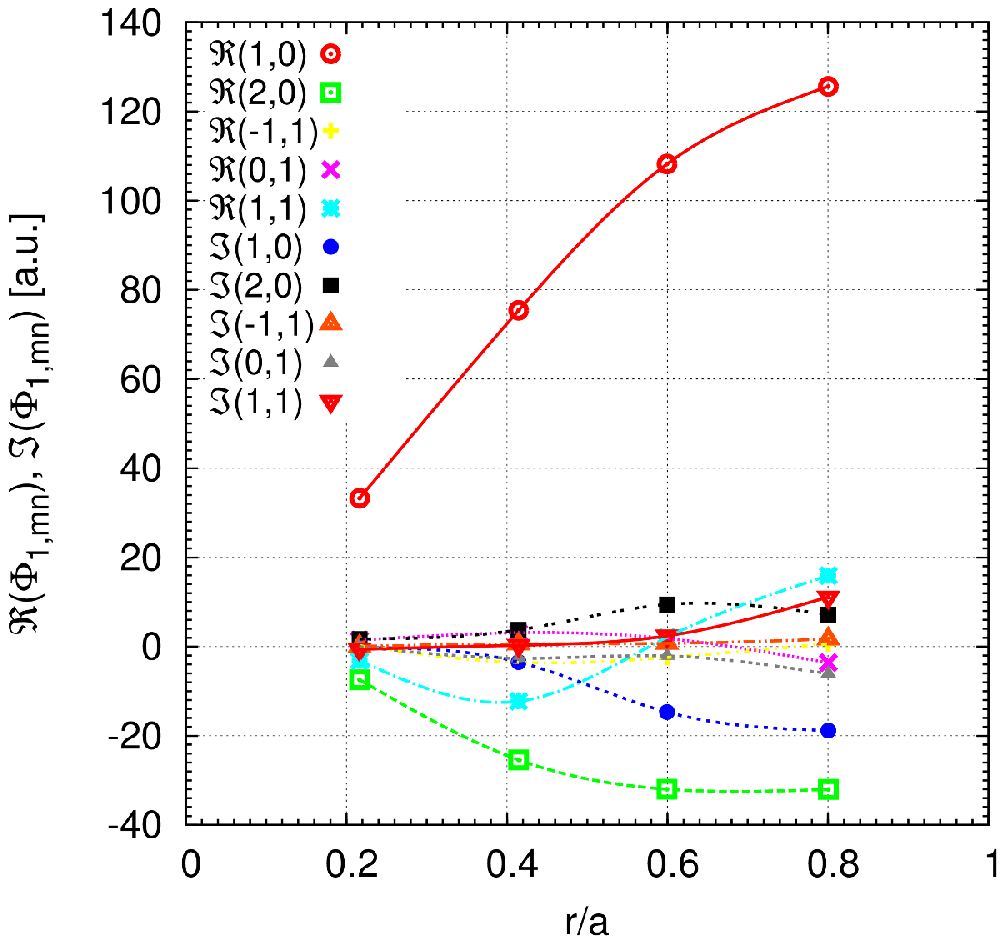}
  \caption{Amplitude of the main Fourier components as a function of the radial position for the 
six cases considered, displayed as in figs. \ref{fig:lhd_pflux}.}
  \label{fig:lhd_fourier}
\end{center}
\end{figure}

In order to sketch qualitatively what underlies the fact that the 
radial flux predicted by the standard
neoclassical prediction becomes stronger or weaker
when $\Phi_{1}$ is taken into account, in fig. \ref{fig:lhd_fourier} 
the real and imaginary part of the complex
Fourier coefficients of $\Phi_{1}$, or equivalently 
the amplitude of cosine and sine components, is presented at
each position simulated in the set of figs. \ref{fig:lhd_pflux}. Each of the plots in 
figs. \ref{fig:lhd_fourier} refers to that at the same position 
in figs. \ref{fig:lhd_pflux}.
All the cases point out to the fact that the main components are the $\cos\theta$ and $\sin\theta$, 
and that a correlation between the sign of the latter, plus/minus, corresponds
respectively to the situation where less/ more inwards flux is driven.
This is remarkably clear in the low density temperature scan. Looking at the 
three plots at the bottom of the figures \ref{fig:lhd_pflux} and the 
corresponding Fourier coefficient at the bottom row of figs. \ref{fig:lhd_fourier}, 
the transition from a mitigated to an enhanced inwards flux accurs approximately at the radial position 
where the sign of the $\sin\theta$ component flips the sign.\\

It is important, in oder to provide an idea of the more complex coupling of $\nabla\Phi_1$ 
(in comparison with $\nabla B$) to $f_{1}$ and produce transport, 
the presence of \textit{sine} components in the spectrum of $\Phi_{1}$. This point out to the 
lack of the stellarator symmetry that for instance 
the magnetic field modulus $B$ has. It is indeed the stellarator symmetry of $B$ 
what makes a necessary condition its absence in $\Phi_1$. 
Considering the radial flux density of particles expressed as 

\begin{equation}
\label{eq:pfluxdef}
\left<\boldsymbol{\Gamma}\cdot\nabla r\right>=\left<\int\text{d}^{3}v \frac{\text{d} r}{\text{d} t}f_{1} \right>,
\end{equation}

where $\left<...\right>$ is the flux surface average. For the bulk ion neoclassical flux 
it can be written as $\text{d} r/\text{d} t \approx \mathbf{v}_{\text{d}}\cdot\nabla r$. 
This term introduces derivatives respect to $\theta$ and $\phi$ in the 
integrand of \ref{eq:pfluxdef} that makes
that a stellarator symmetric magnetic field does not drive any transport unless
the distribution function indeed develops \textit{sine} components. This sine components
are then inherited by the moments of the perturbed part of the distribution function like 
the perturbed density and subsequently the electrostatic potential $\Phi_1$.
Regarding the impurity radial particle transport one needs to recover the initial hypothesis
that retaining $\mathbf{v}_{E1}$ is necessary and express 
$\text{d} r/\text{d} t \approx \left(\mathbf{v}_{\text{d}}+\mathbf{v}_{E1}\right)\cdot\nabla r$.
\blue{Then, regarding how $\Phi_{1}$ participates on the transport of impurities becomes less trivial
than for the case of $B$, given that $\Phi_1$ Fourier spectrum has both \textit{sine} and \textit{cosine} parts 
that drives particle flux coupled to the perturbed distribution function through both 
its \textit{cosine} and \textit{sine} components respectively, and not only through the latter ones as $\mathbf{v}_{d}$ does.
To this respect the question of how efficient can a specific component of $\Phi_1$ be 
in counteracting the trend of the impurities to accumulate depends on the spatial 
dependence of their distribution function, and brings to the same importance the 
the knowledge of $\Phi_1$ and the impurity density variation on the flux surface. 
An indeed, both magnitudes have recently met an increasing interest in their measurement 
\cite{Viezzer_ppcf_55_124037_2013, Arevalo_nf_54_013008_2014, Pedrosa_arXiv_1404.0932_2014}.
is kept out of the scope of this paper will be addressed in future works.}

\subsection{Wendelstein 7-X results}
\label{sec:w7x}

\begin{figure}[t]
\begin{center}
  \includegraphics[width=0.3\textwidth]{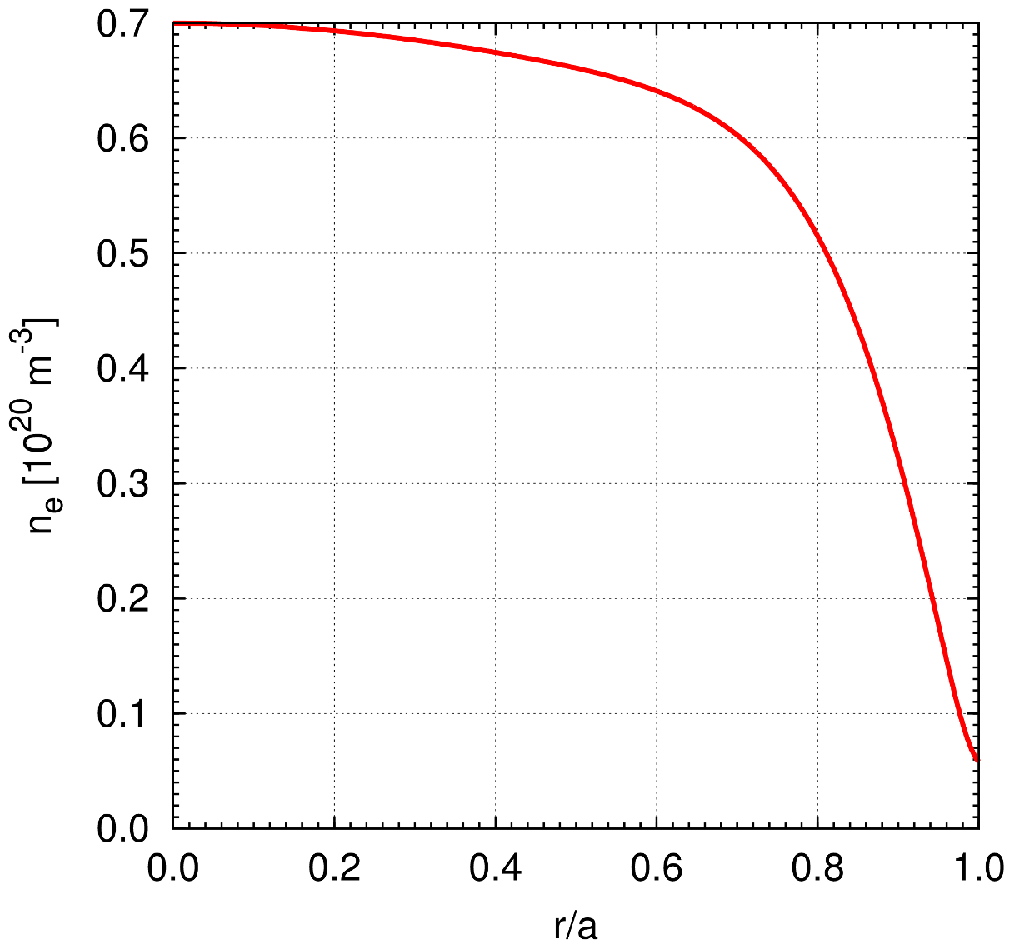}\hspace{0.0cm}
  \includegraphics[width=0.3\textwidth]{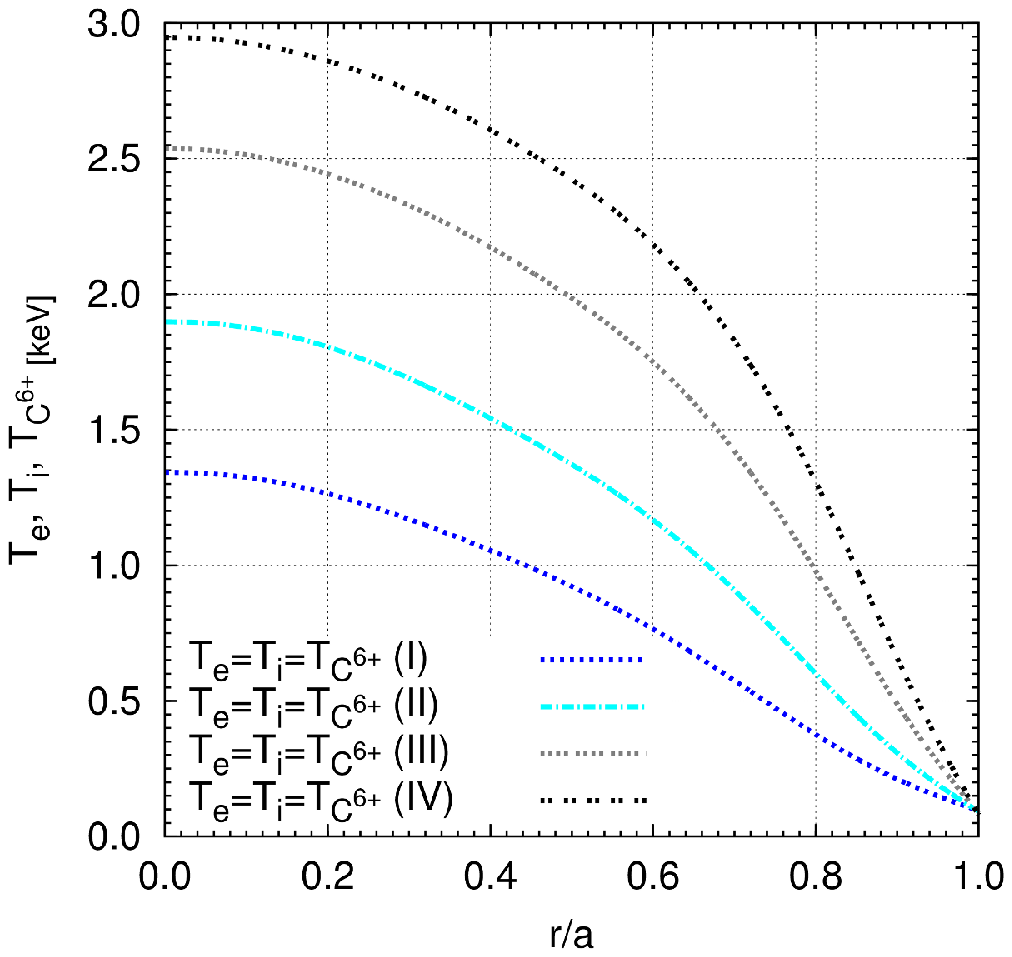}\hspace{0.4cm}
  \includegraphics[width=0.32\textwidth]{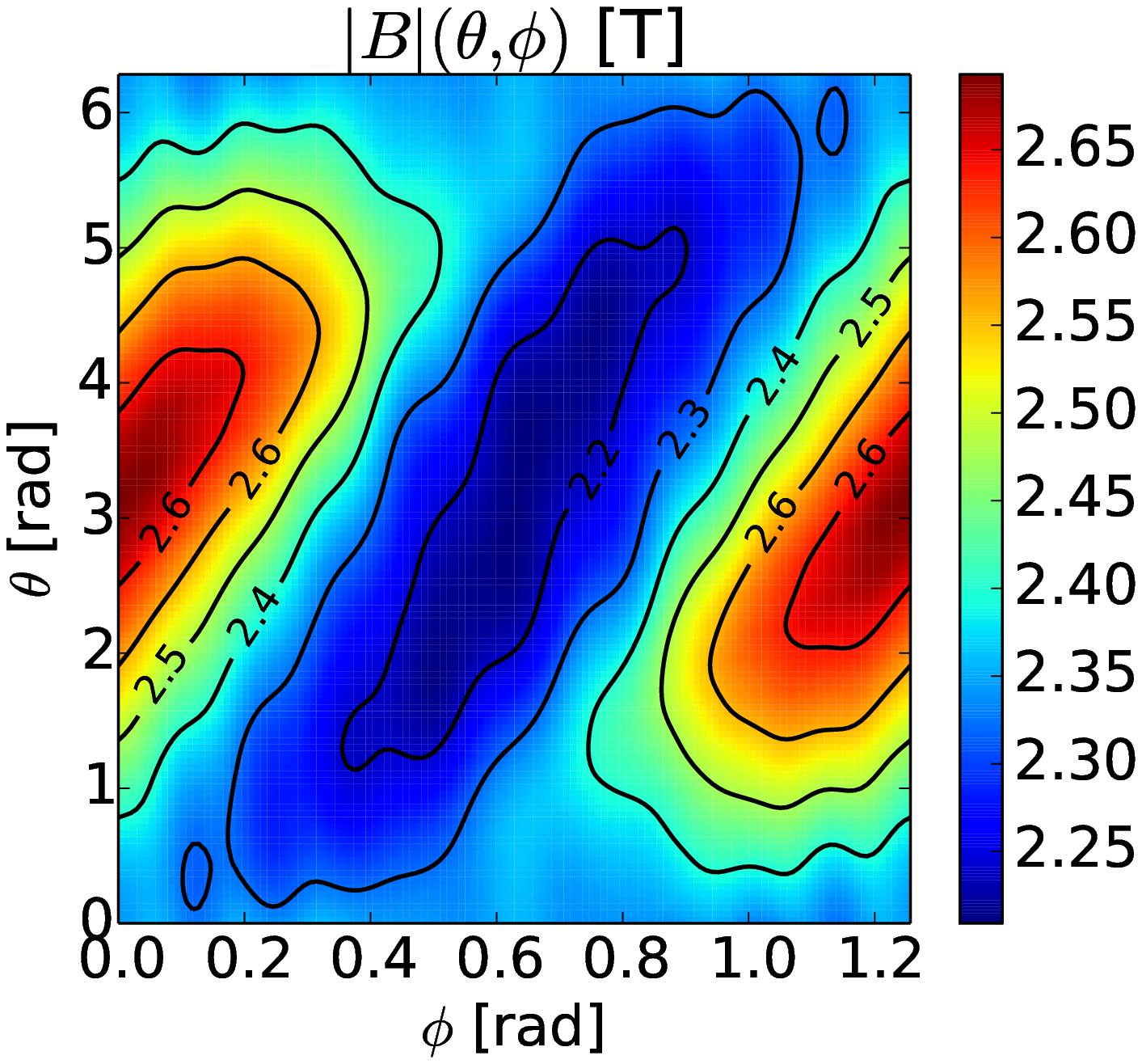}
  \caption{Set of profiles considered for W7-X. (Left) Electron density profile. 
(Center) Electron, bulk ion and C$^{6+}$ temperature profiles, assumed the same for all the three species. 
(Right) Magnetic field modulus in one W7-X period at the position $r/a=0.6$.}
  \label{fig:w7x_profiles}
\end{center}
\end{figure}

The main features of the dependence of $\Phi_{1}$ with the plasma parameters, 
as well as the role of its Fourier spectrum on supporting
or reducing the tendency of C$^{6+}$ to accumulate have been discussed in sec. \ref{sec:lhd}.
In this one we show similar numerical simulations for the Wendelstein 7-X stellarator,
and a set of profiles expected under operation. These are shown in fig. \ref{fig:w7x_profiles}. The electron density
profile is represented in fig. \ref{fig:w7x_profiles} (left), and as before $Z_{\text{eff}}=1.1$
is considered to determine $n_{0i}$ and $n_{0Z}$.
The temperature, the same for all species in each of the two cases,
is the parameter scanned. Four different profiles have been considered.
The temperature profiles in fig. \ref{fig:w7x_profiles} (center) 
are labeled as \text{I},  \text{II}, \text{III} and \textit{IV} increasingly 
with $T$.\\

\begin{figure}
\begin{center}
  \includegraphics[width=0.23\textwidth]{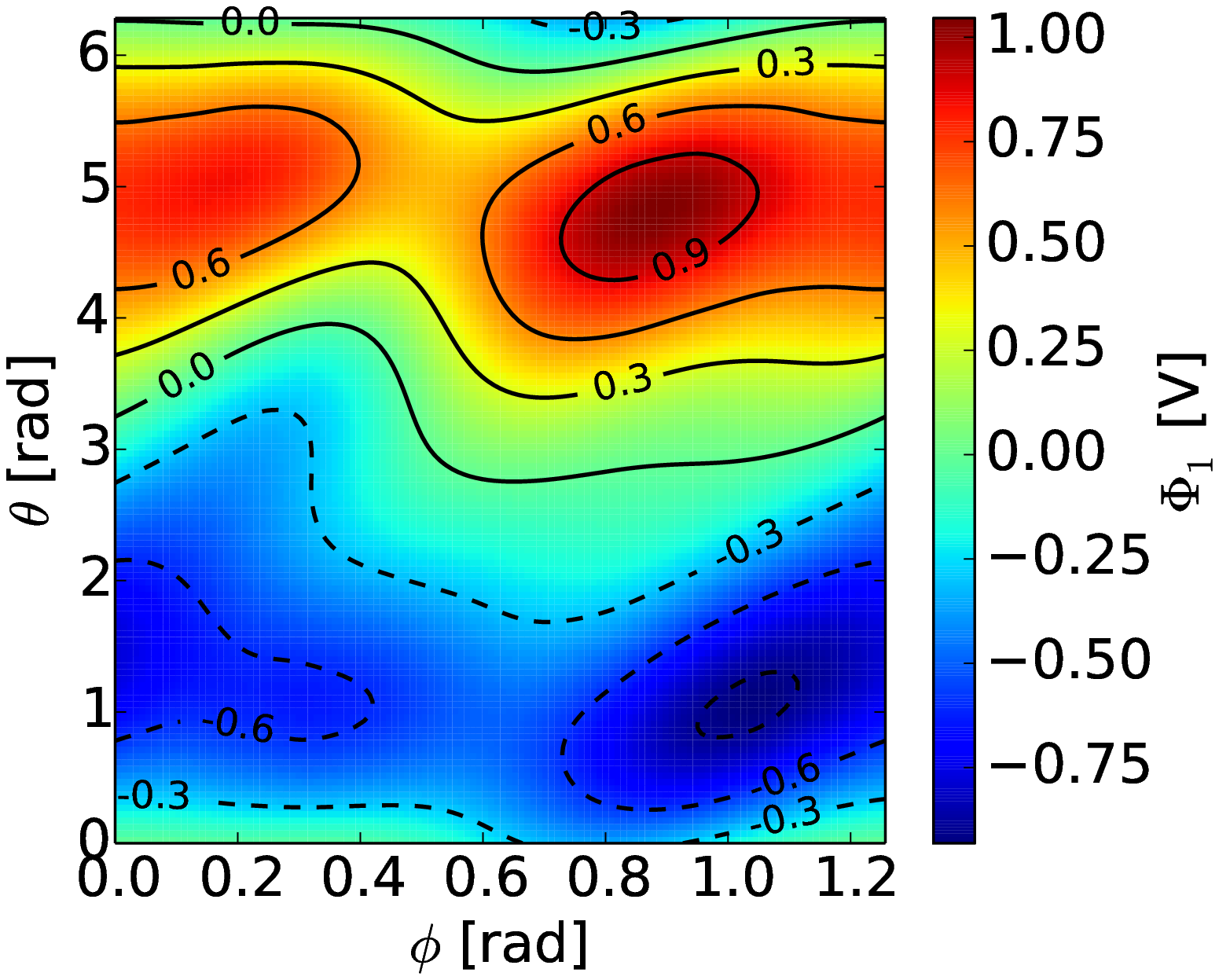}\hspace{0.2cm}
  \includegraphics[width=0.23\textwidth]{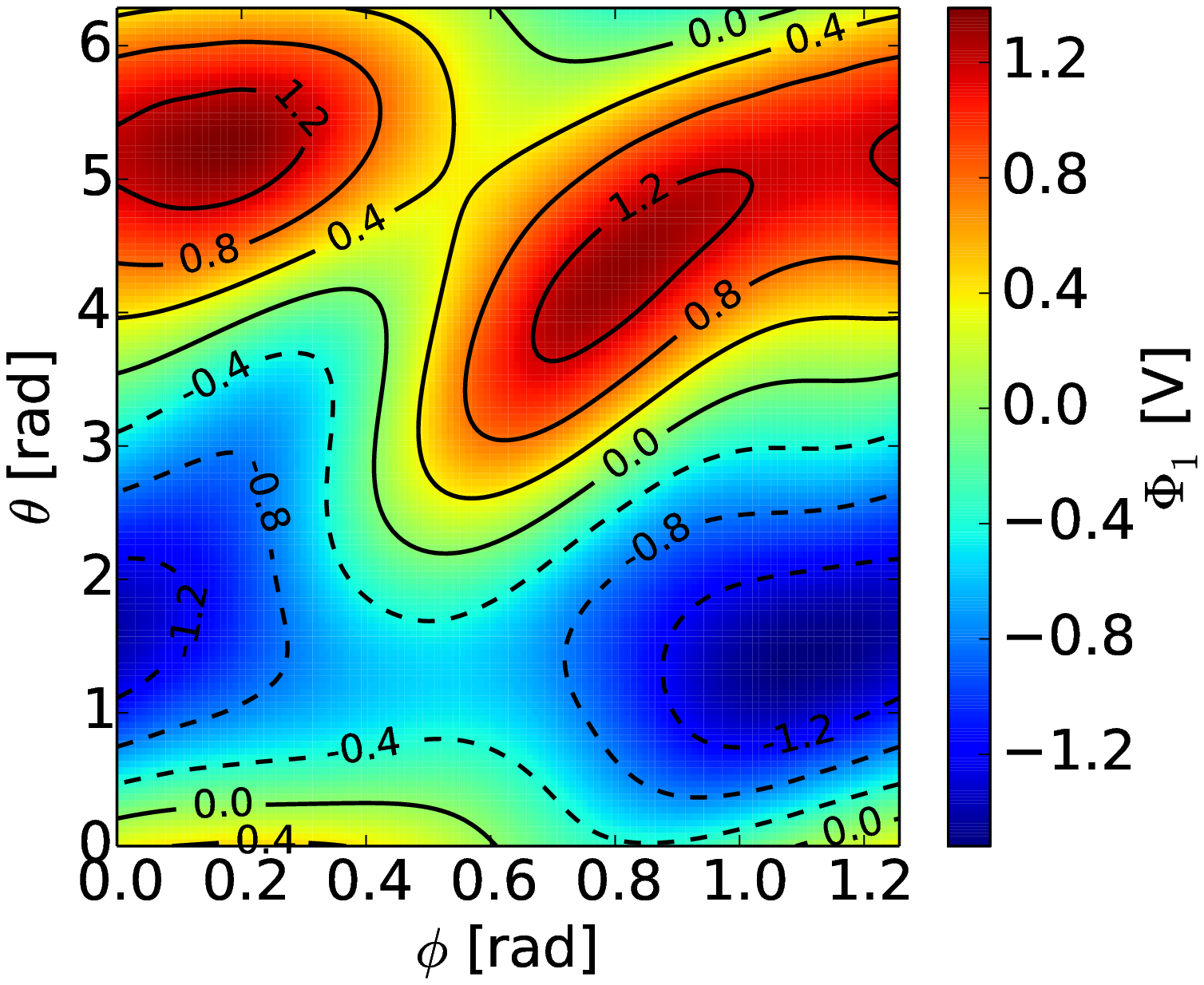}\hspace{0.2cm}
  \includegraphics[width=0.23\textwidth]{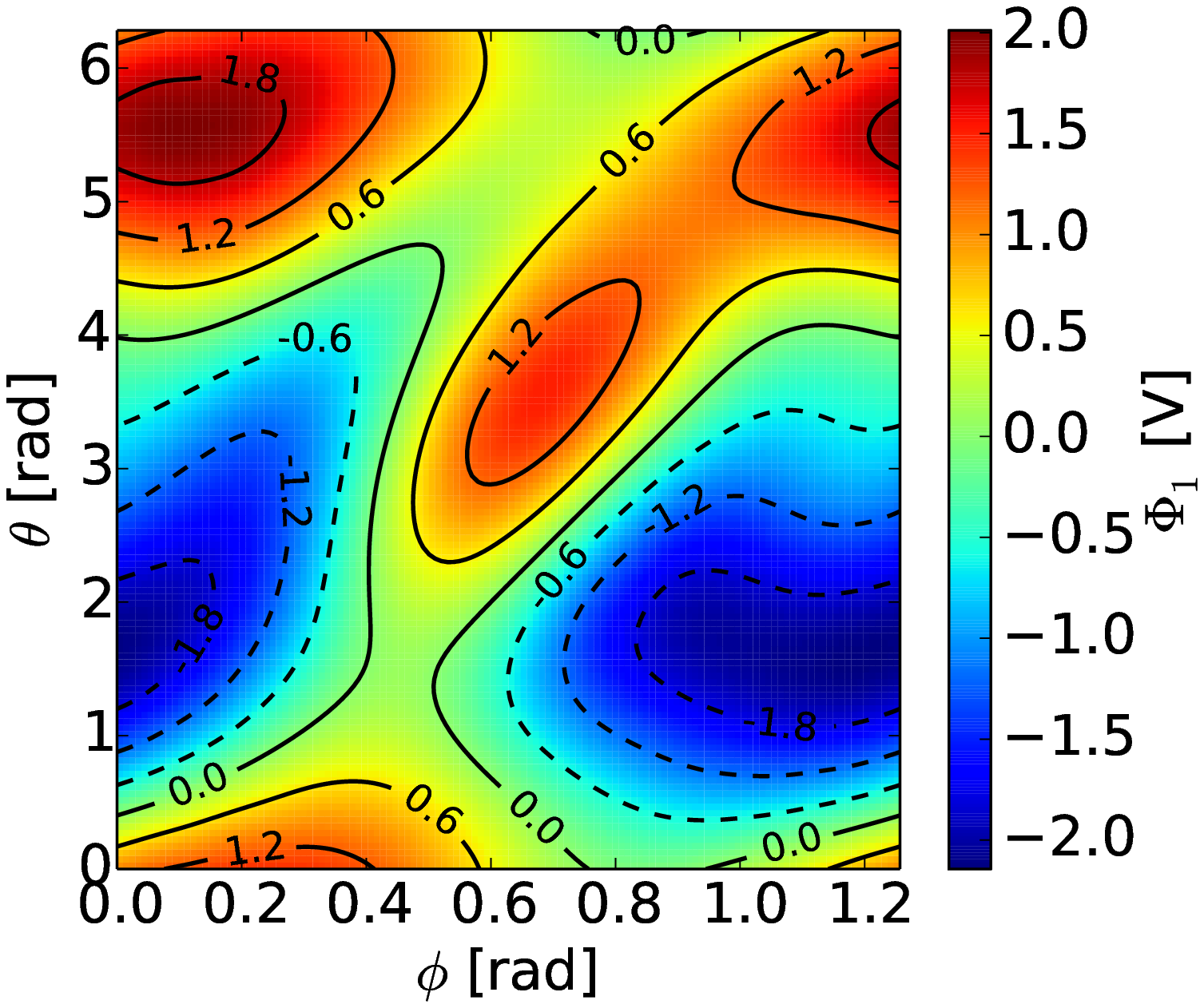}\hspace{0.2cm}
  \includegraphics[width=0.23\textwidth]{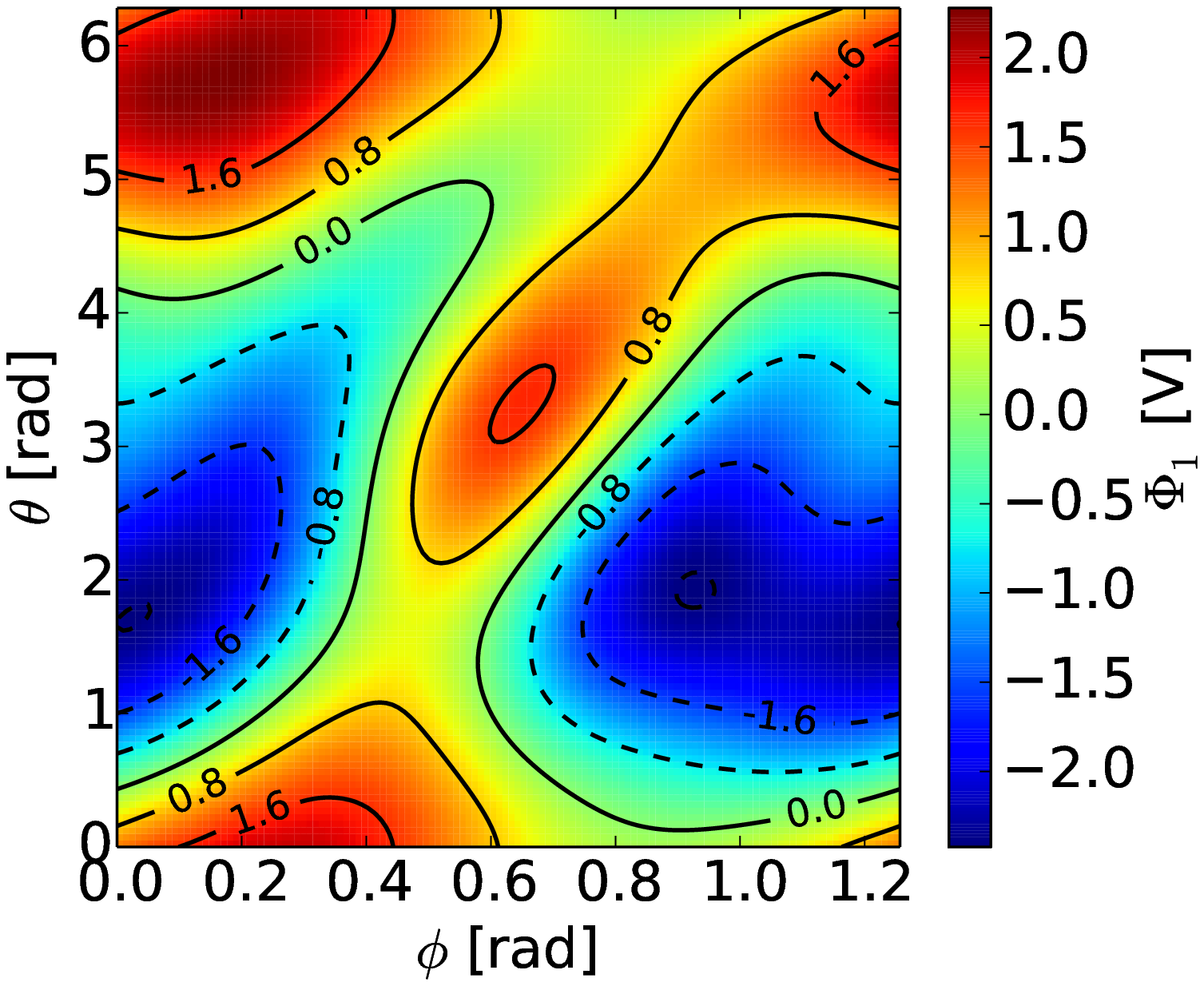}\\
  \includegraphics[width=0.23\textwidth]{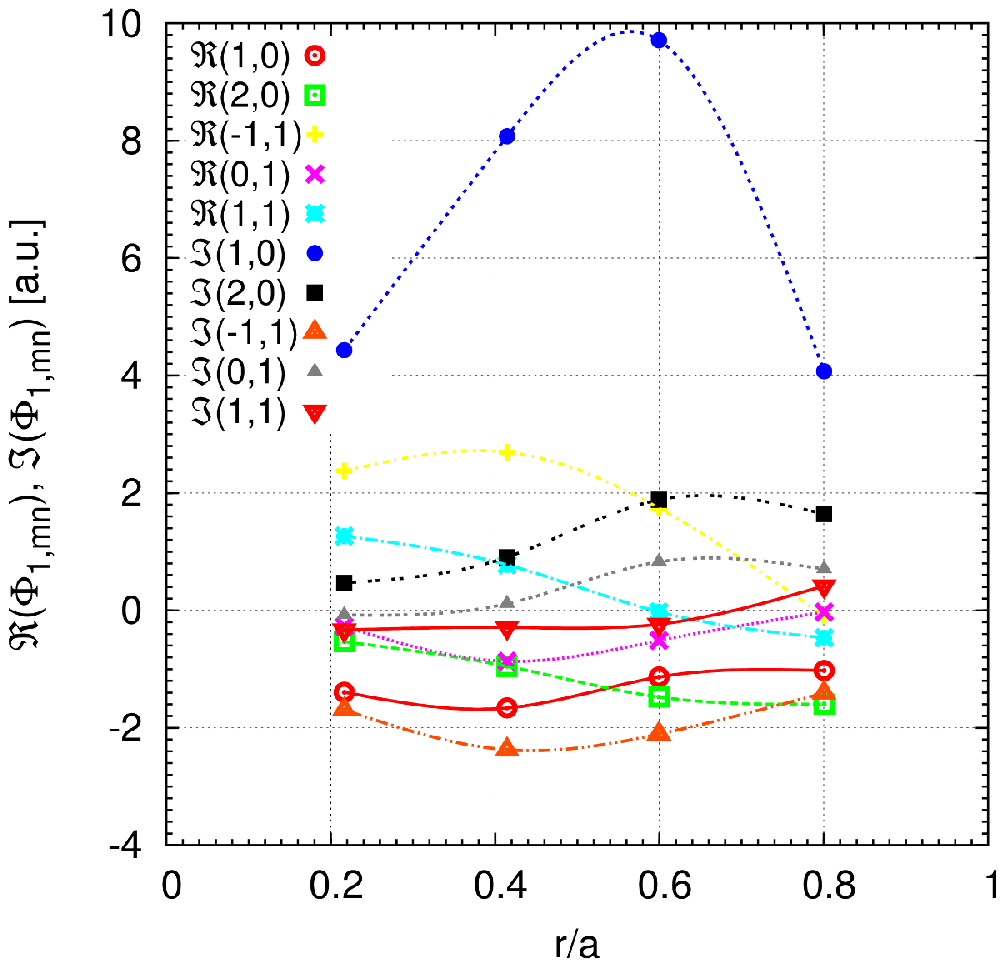}
  \includegraphics[width=0.23\textwidth]{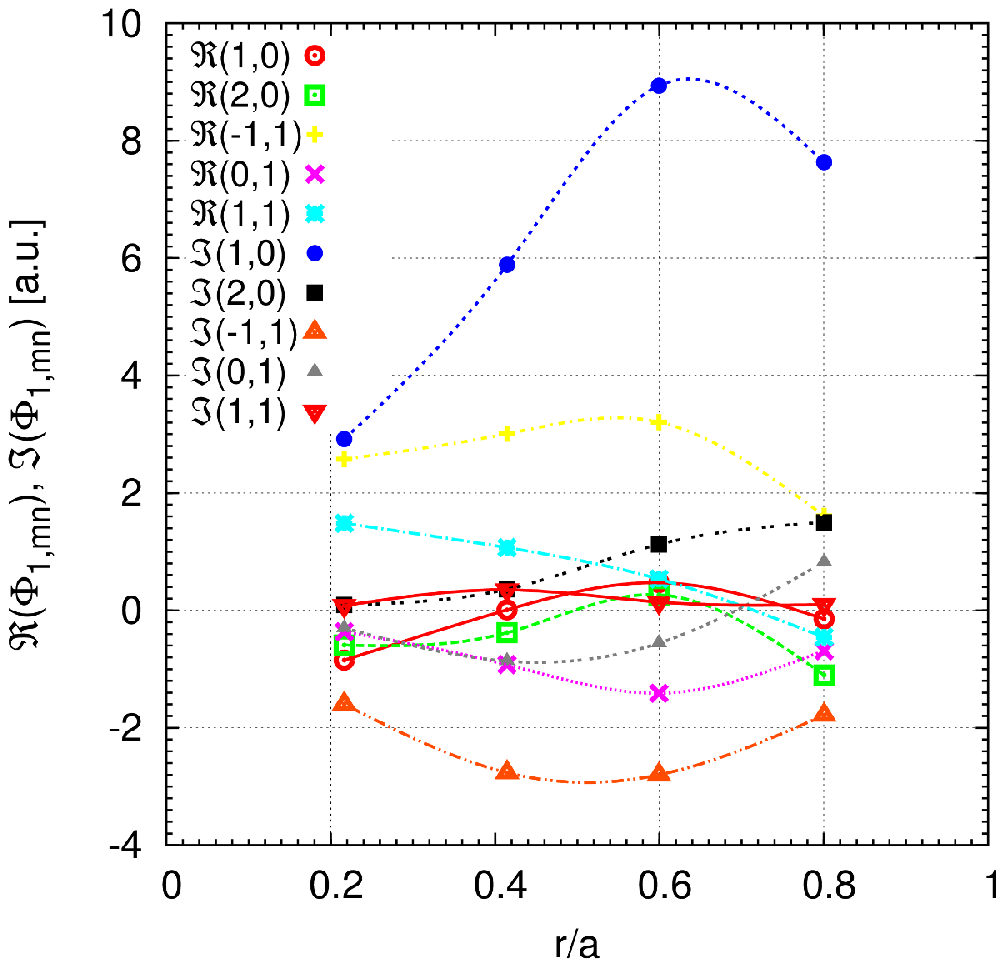}
  \includegraphics[width=0.23\textwidth]{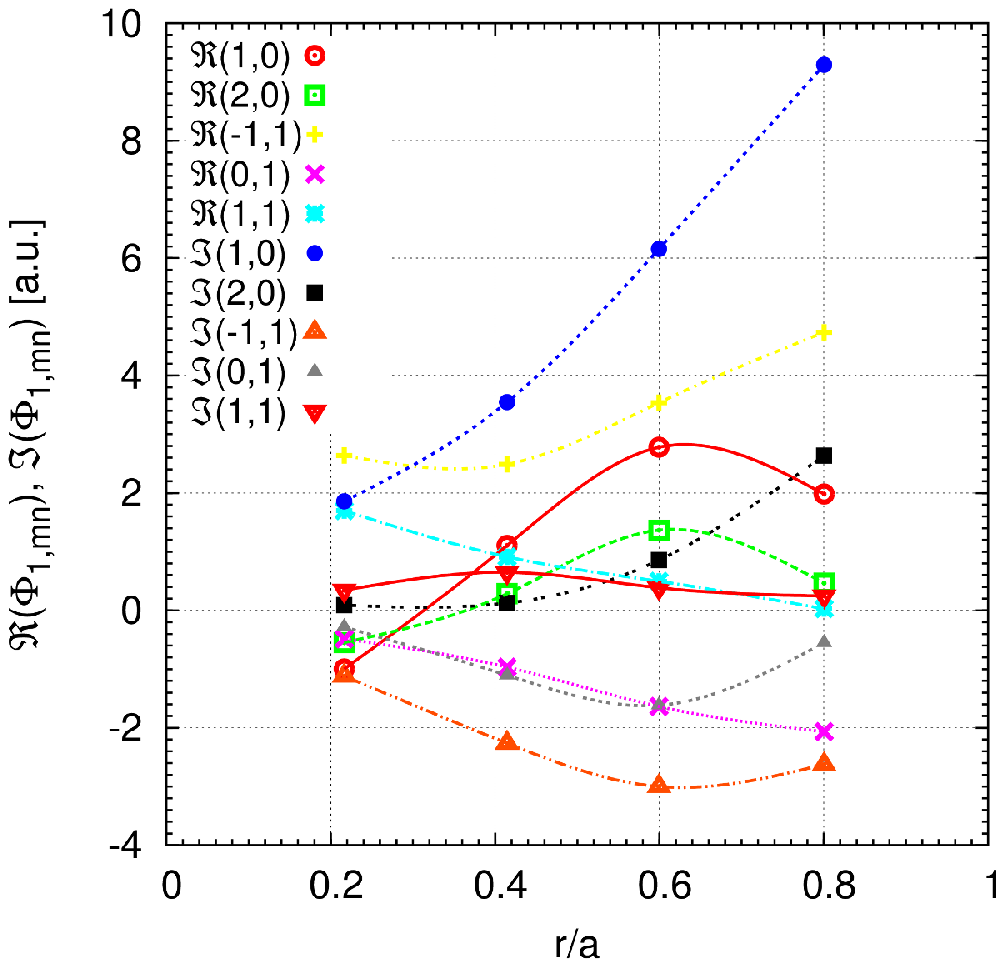}
  \includegraphics[width=0.23\textwidth]{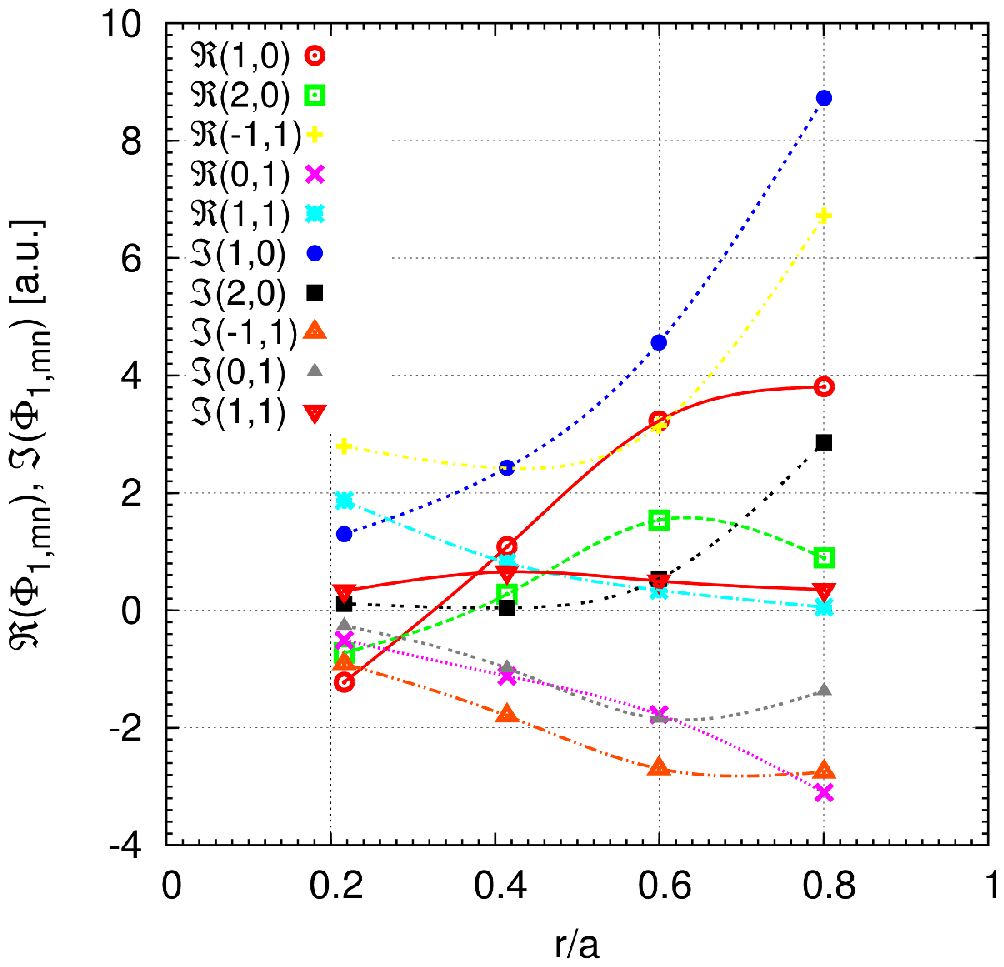}
  \caption{At the position $r/a=0.6$ for W7-X, $\Phi_{1}(\theta,\phi)$ for the temperature
profiles \textit{I}, \textit{II},\textit{III} and \textit{IV} (from left to right).}
  \label{fig:w7x1_phi1maps}
\end{center}
\end{figure}

The electrostatic potential is in this case appreciably much lower 
than at LHD. Although the collisionality does not 
reach in any of the 
the four cases considered in this section the value of the lowest collisional low density 
LHD set $\textit{B}$, the variation of the potential 
is noticeably much lower in W7-X for a similar collisionality.\\

\blue{A possible reason underlying the low electrostatic potential
variation in W7-X is the physical target for the design of this device, namely, reducing the 
neoclassical losses and bootstrap current by approaching an omnigeneous magnetic field structure. 
In a perfect omnigeneous magnetic field the bounced-averaged 
drift orbits of the localized particles align to the flux surfaces. 
Although this ideal situation is not, and cannot be \cite{Garren_pfb_3_2822_1991}, the case, the closer the W7-X magnetic field structure
is from it -- compared to the other devices considered in this work -- certainly leads to a
smaller departure of the localized particles from the flux surfaces
and results in a weaker electrostatic potential variation on them.}
Regarding the Fourier spectrum of $\Phi_1$, similar 
features to LHD's at similar collisionality (cases A.I-III) are shown. A 
dominant $\sin\theta$ component, and a weaker $\cos\theta$ one
that increases as collisionalily decreases is the footprint of the electrostatic potential in both $W7-X$
and LHD, although the rest of the modes represent a more noticeable contribution 
to the spectrum of $\Phi_1$ than in LHD, where they are all marginally zero with 
a some exception (e.g. the $\cos 2\theta$).
  
\begin{figure}[t]
\begin{center}
  \includegraphics[width=0.23\textwidth]{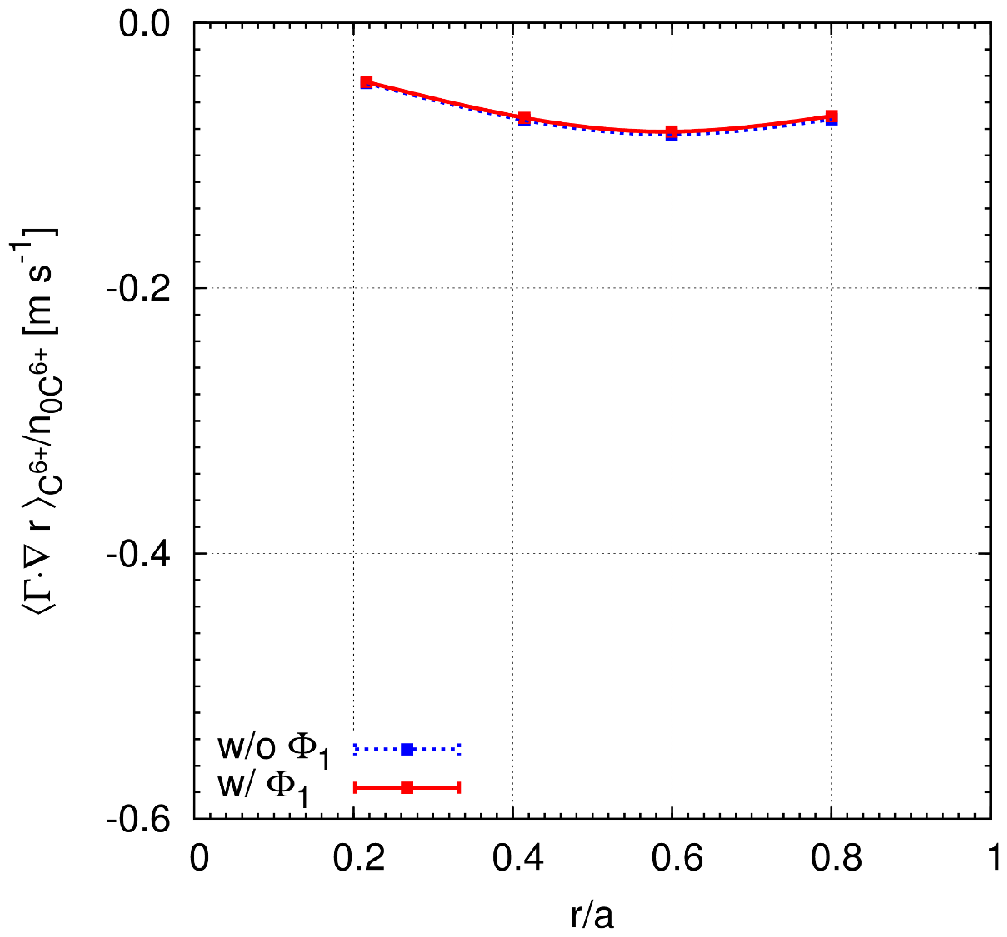}
  \includegraphics[width=0.23\textwidth]{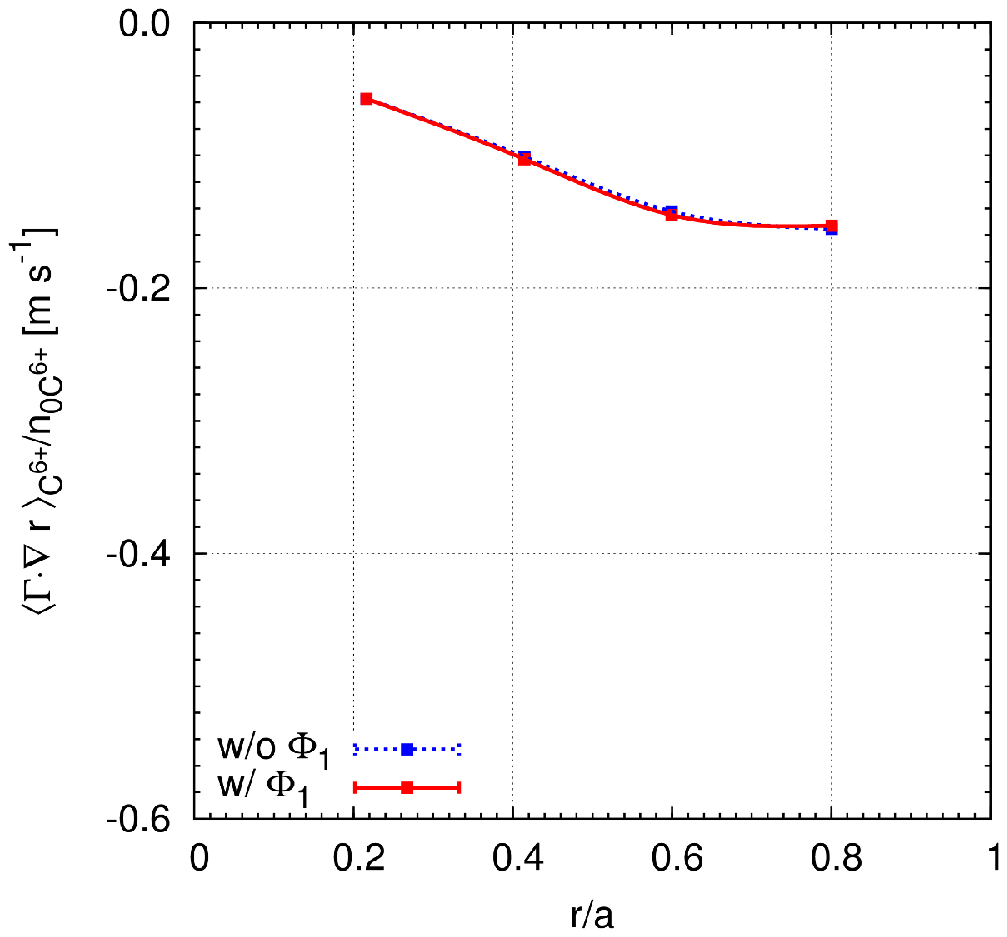}
  \includegraphics[width=0.23\textwidth]{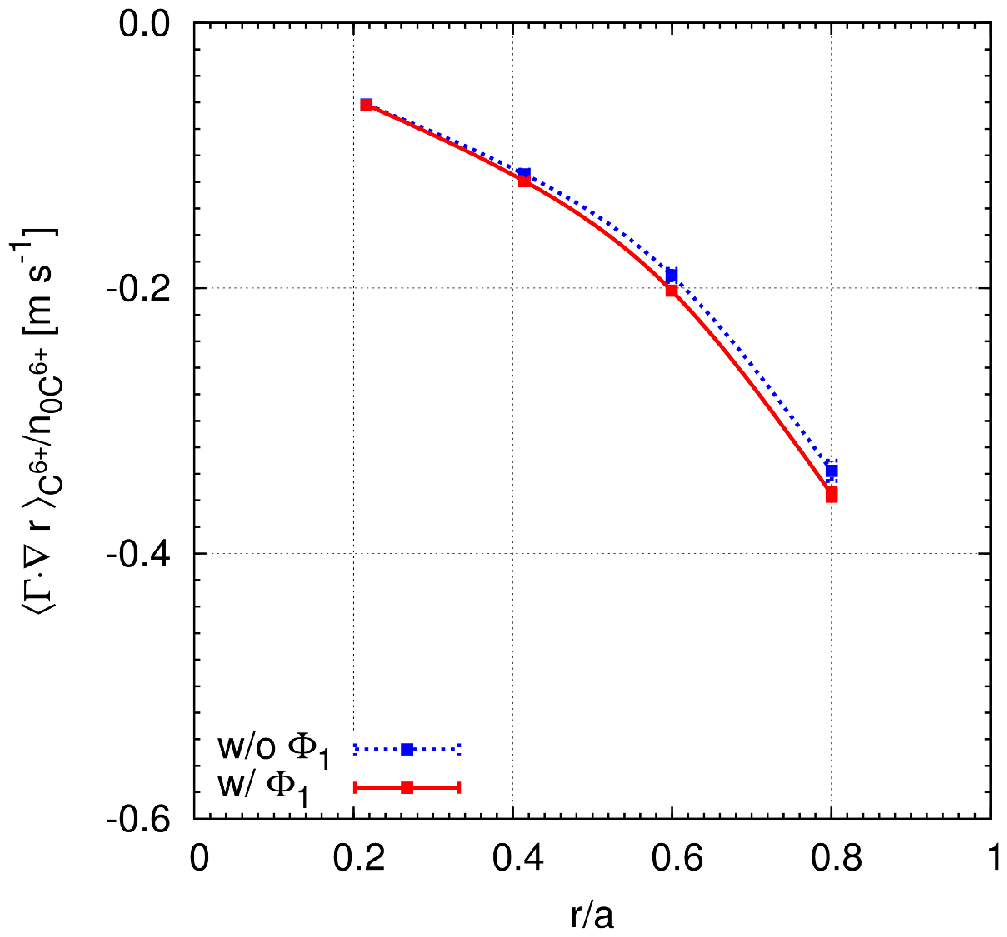}
  \includegraphics[width=0.23\textwidth]{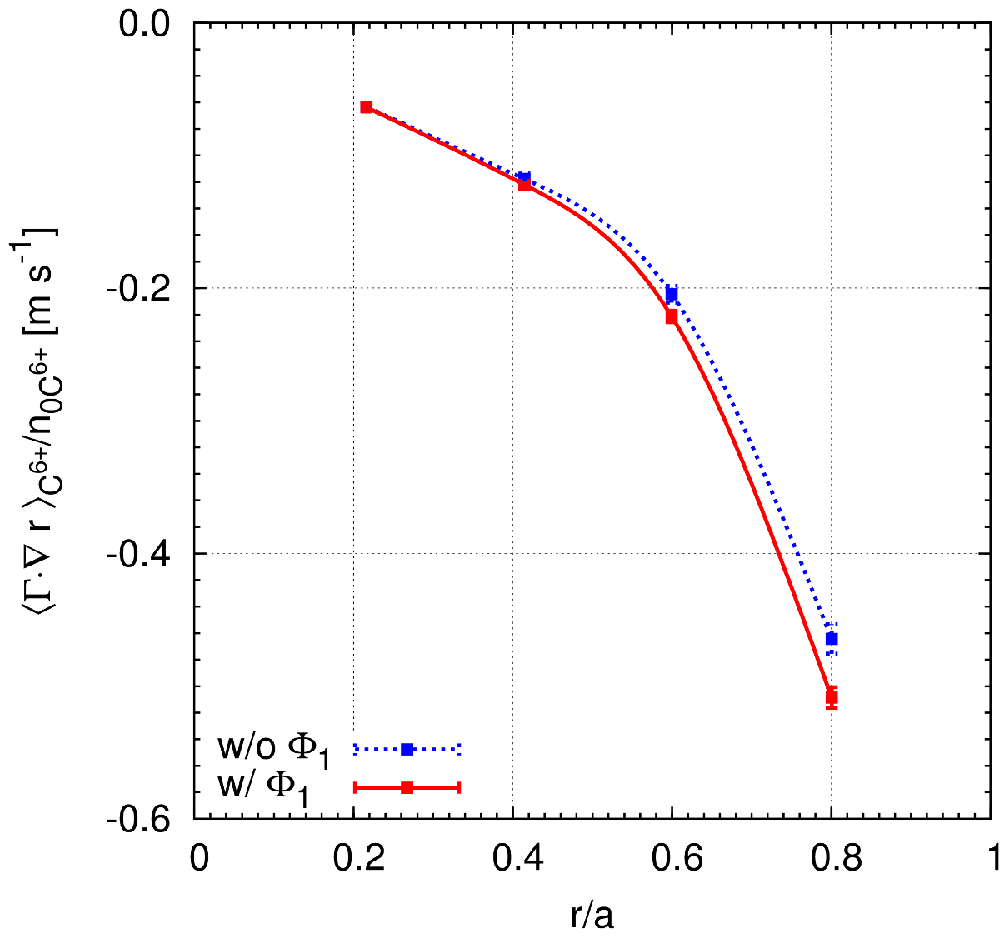}\\
  \caption{Radial flux density of C$^{6+}$ as a function of $r/a$ including $\Phi_{1}$ (solid line) and 
neglecting it (dotted line). The cases \textit{I} to \textit{IV} are represented from left to right.}
  \label{fig:w7x1_pflux}
\end{center}
\end{figure}

The radial profiles for the particle flux density of C$^{6+}$ are represented
in fig. \ref{fig:w7x1_pflux} in the three W7X cases, 
increasing the temperature from the left to the right. Not surprisingly the difference that $\Phi_{1}$ 
makes respect to the standard neoclassical prediction can be well considered for the 
present set of parameters and impurity species negligible and aligned to the neoclassical 
optimization aforementioned.

\subsection{TJ-II results}
\label{sec:tj20}

\begin{figure}[t]
\begin{center}
  \includegraphics[width=0.3\textwidth]{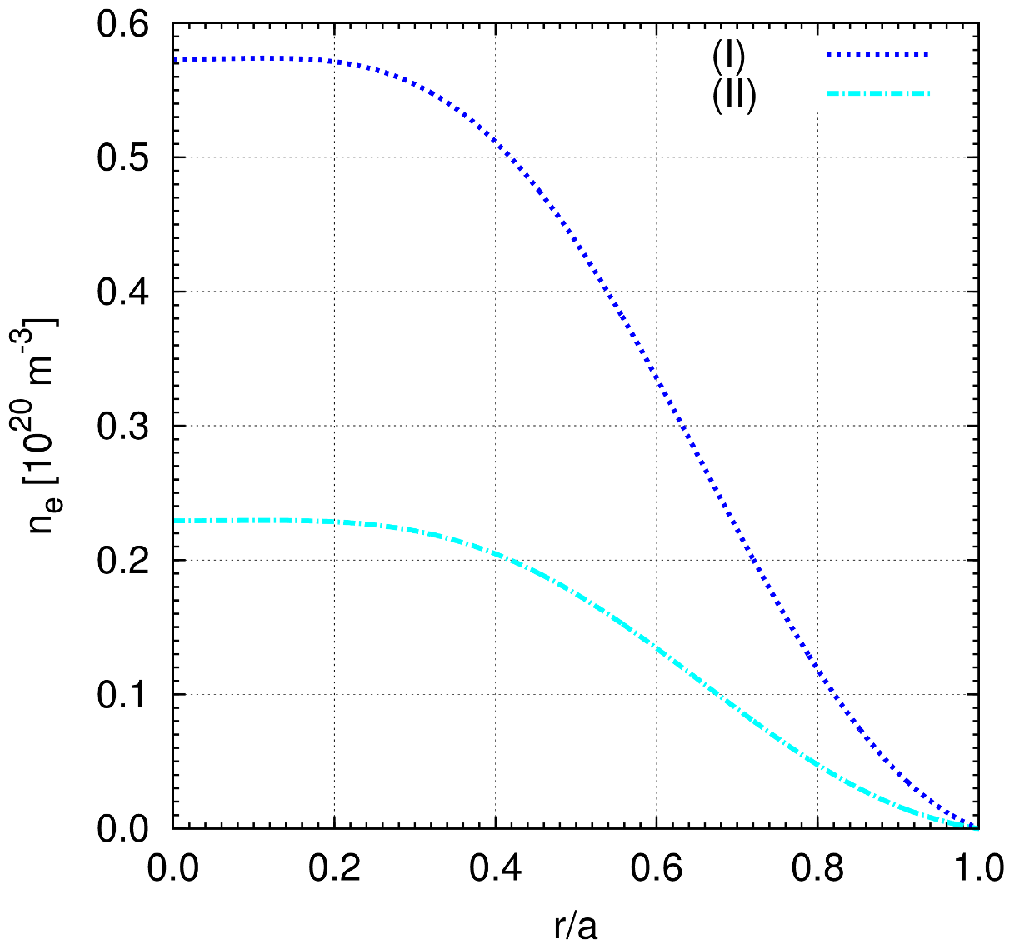}
  \includegraphics[width=0.3\textwidth]{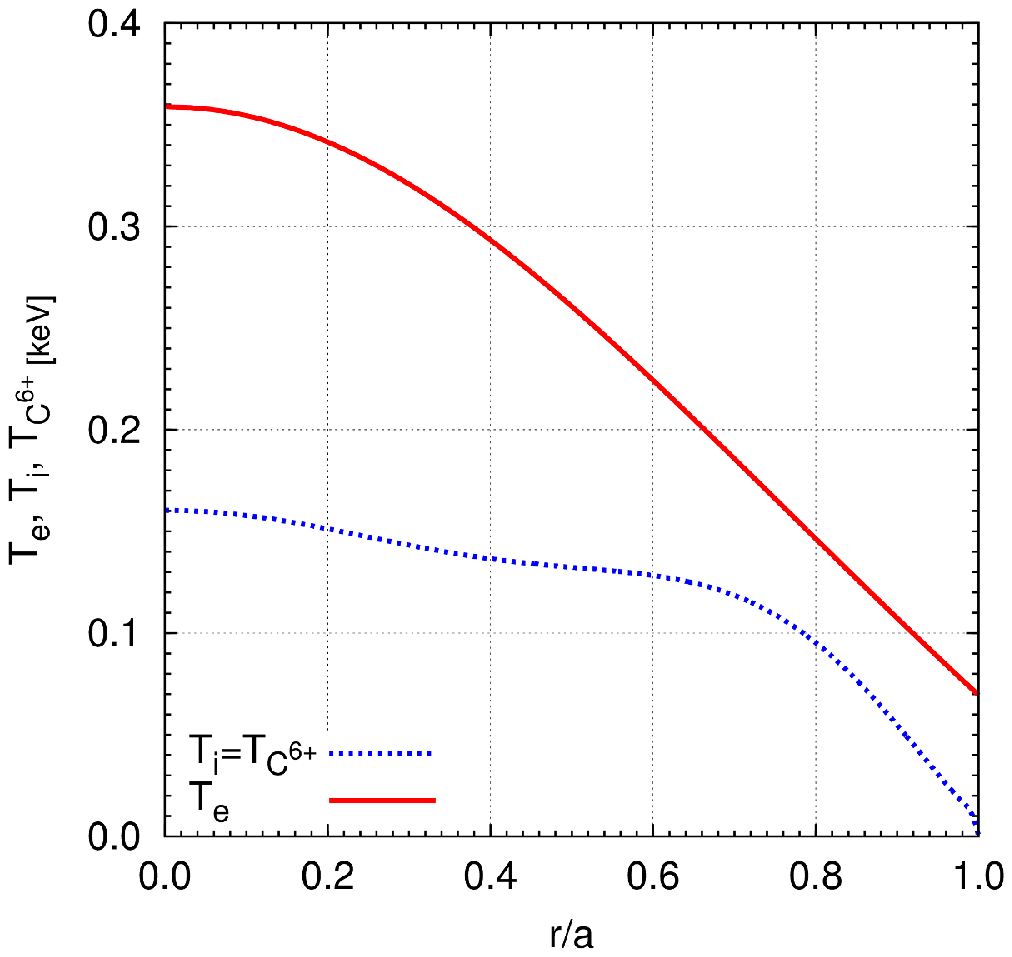}
  \includegraphics[width=0.32\textwidth]{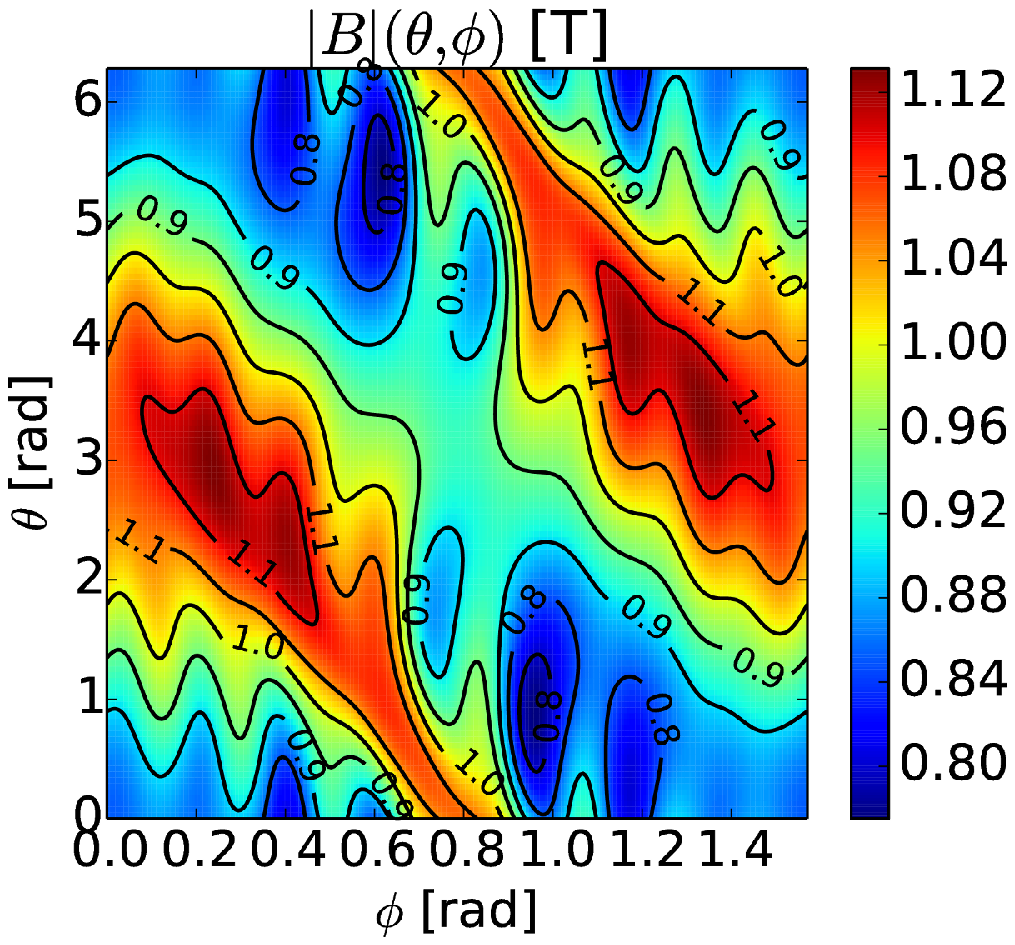}
  \caption{Set of profiles considered for TJ-II. (Left) Electron density radial profile. 
(Center) Electron, bulk ion and C$^{6+}$ temperature profiles, assumed all equal. 
(Right) Magnetic field modulus in one TJ-II period at the position $r/a=0.6$.}
  \label{fig:tjii_profiles}
\end{center}
\end{figure}

Finally in the present section two cases are considered for the TJ-II stellarator.
Two density profiles, shown in fig. \ref{fig:tjii_profiles}(left), at a fixed temperature for the 
electrons and bulk and impurity ions ($H^{+}$ and C$^{6+}$ as in the previous cases)
represented in fig. \ref{fig:tjii_profiles}(center). These 
profiles correspond to typical NBI-heated TJ-II plasmas.
The contour plot of the magnetic field at the flux surface $r/a=0.6$ is represented 
in fig. \ref{fig:tjii_profiles}(right).

\begin{figure}
\begin{center}
  \includegraphics[width=0.3\textwidth]{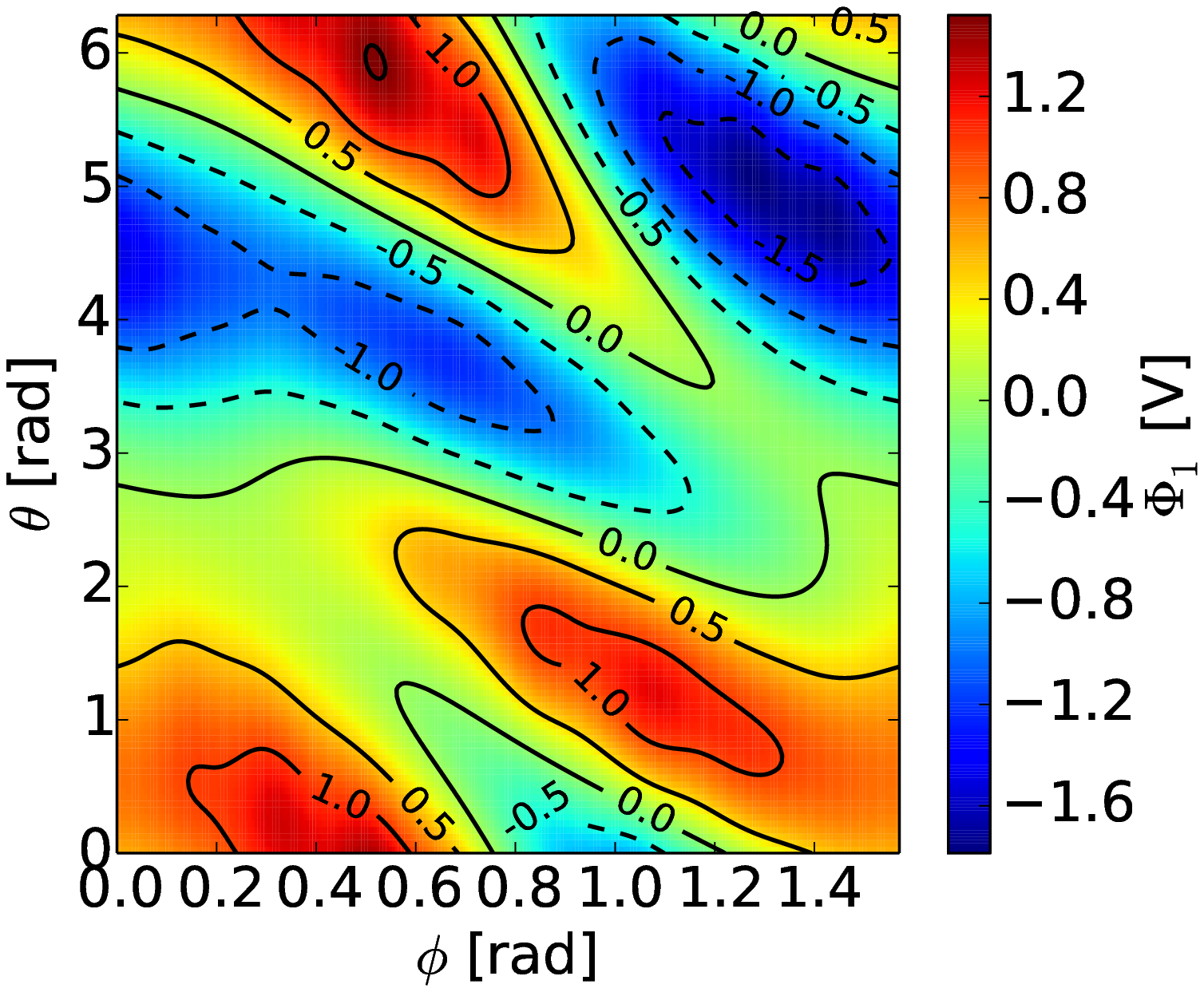}
  \includegraphics[width=0.3\textwidth]{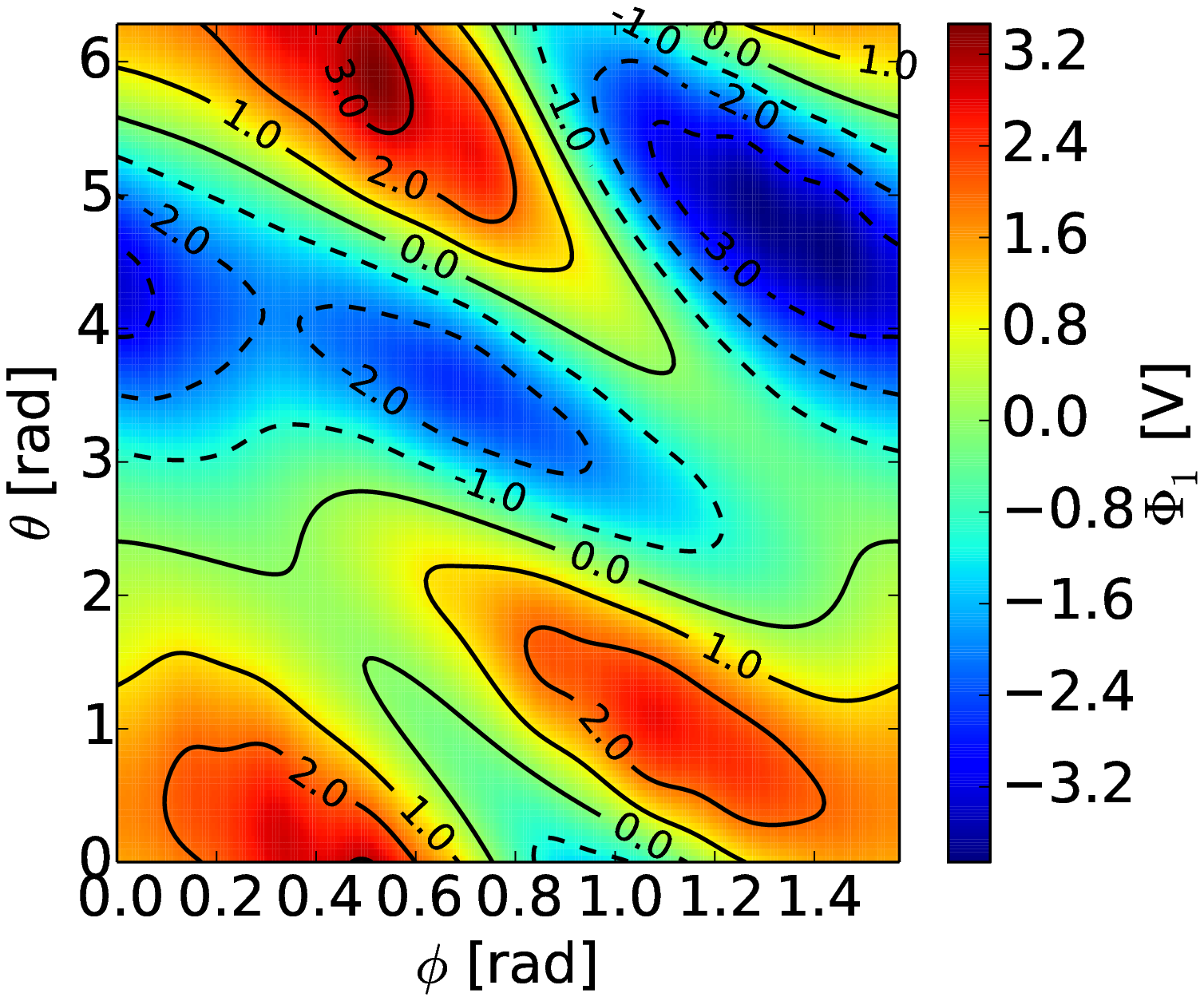}\\
  \includegraphics[width=0.3\textwidth]{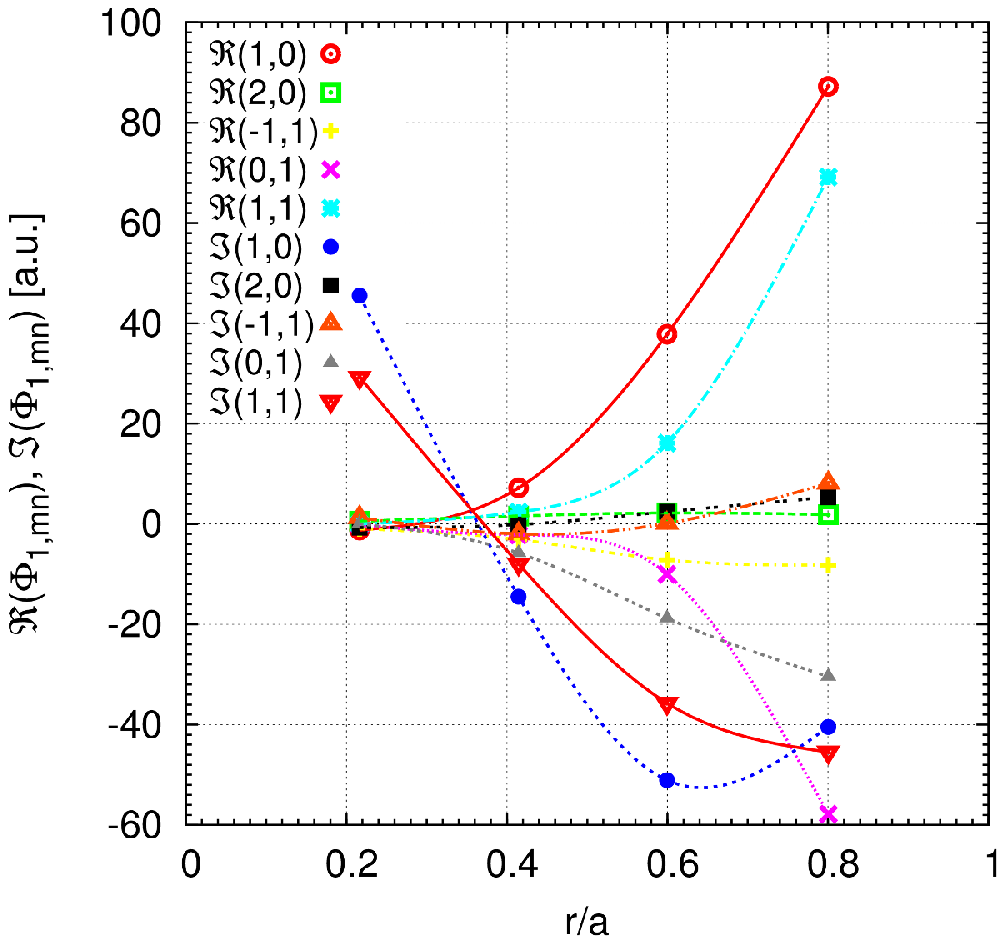}
  \includegraphics[width=0.3\textwidth]{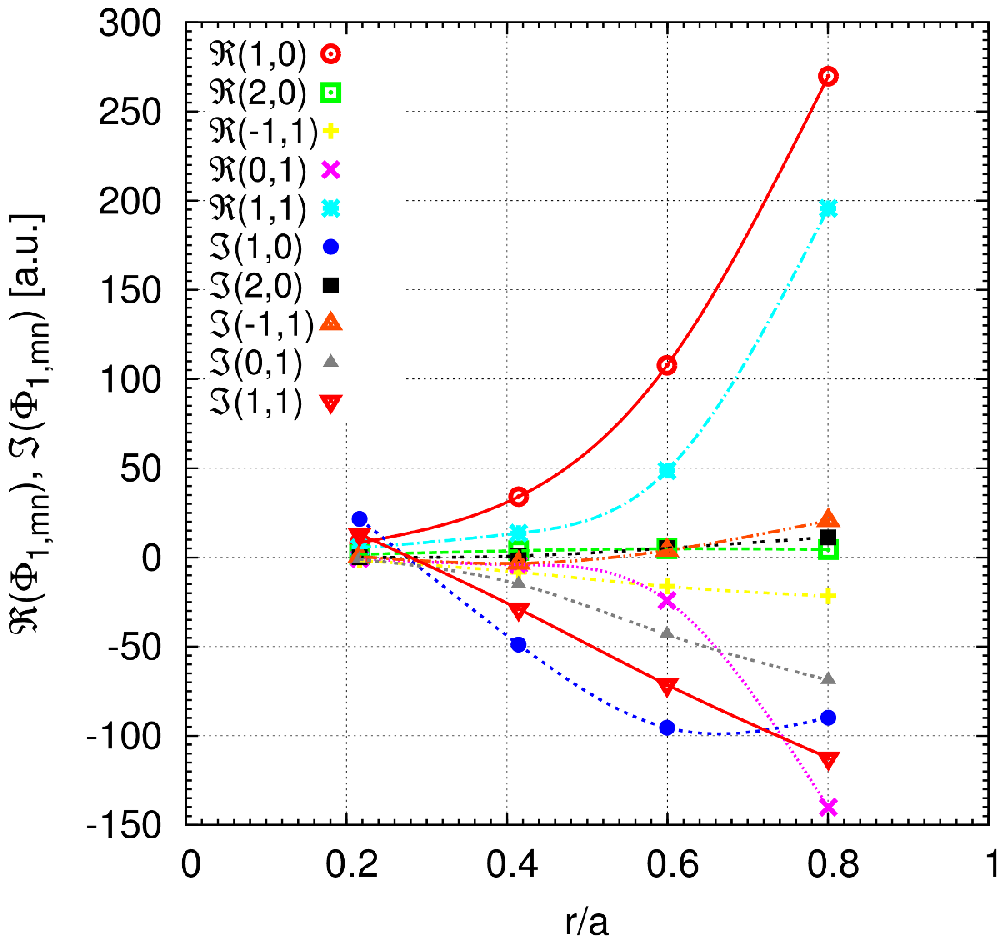}\\  
  \caption{At the position $r/a=0.6$ for TJ-II standard configuration: $\Phi_{1}(\theta,\phi)$ for the temperature
profiles I (left) and II (right).}
  \label{fig:tj20_phi1maps}
\end{center}
\end{figure}

\begin{figure}[t]
\begin{center}
  \includegraphics[width=0.23\textwidth]{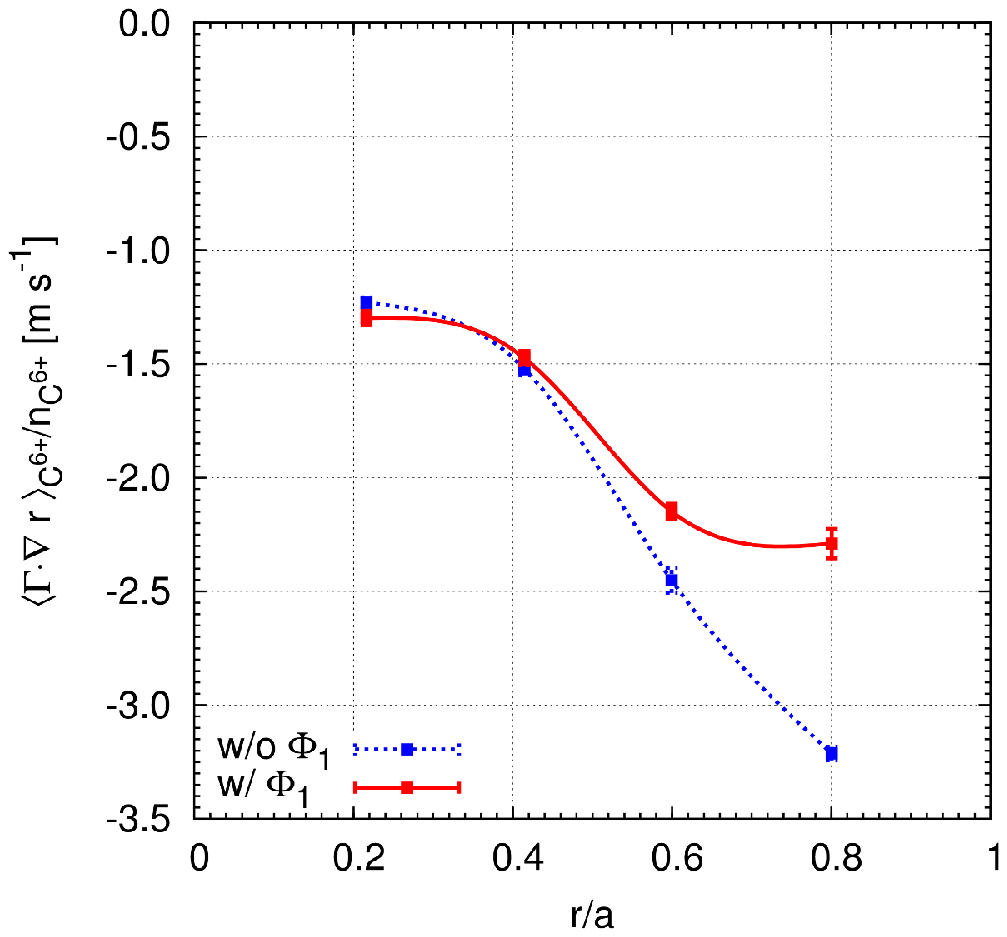}
  \includegraphics[width=0.23\textwidth]{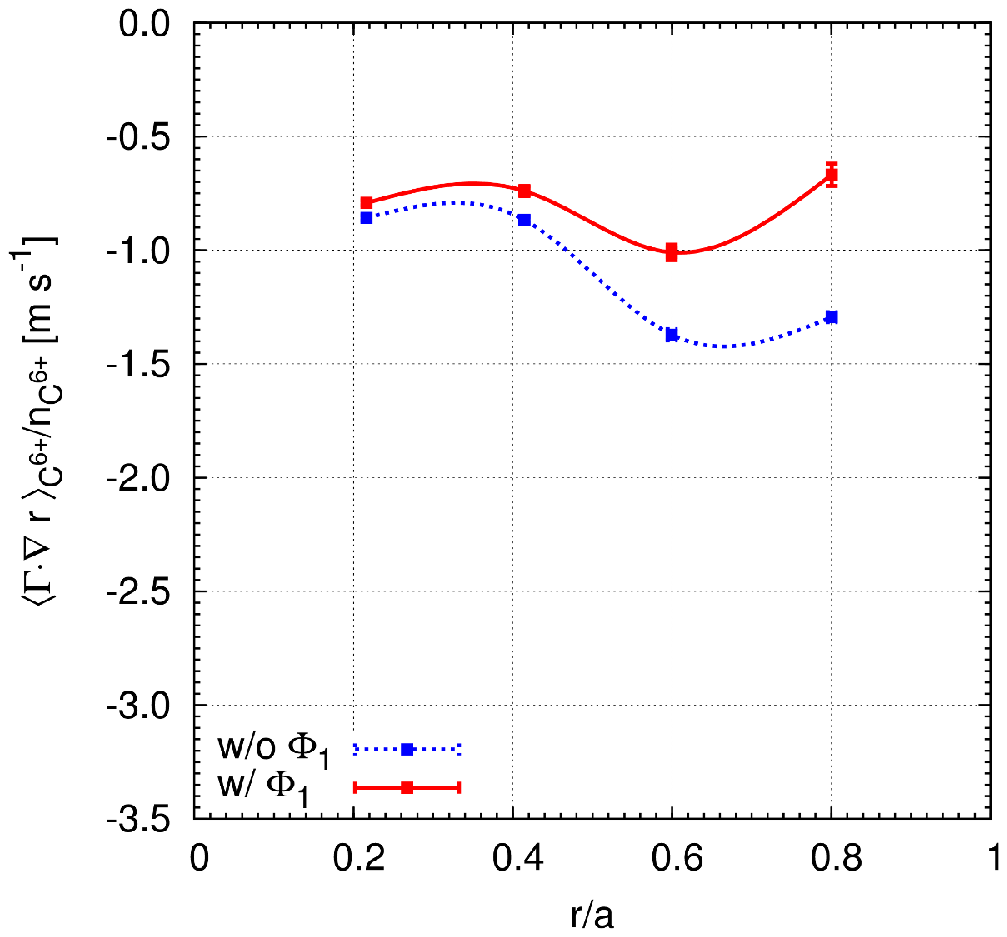}
  \caption{Radial particle flux density of C$^{6+}$ as a function of $r/a$ including $\Phi_{1}$ (solid line) and 
neglecting it (dotted line), for the high and low density cases (left and right respectively).}
  \label{fig:tj20_pflux}
\end{center}
\end{figure}

In fig. \ref{fig:tj20_phi1maps} (top) the electrostatic potential $\Phi_1$
is represented for the two cases considered together with some Fourier harmonics (bottom).
The calculation of $\Phi_1$ performed shows a low $\Phi_1$ amplitude
compared to the LHD cases, and of the same order of magnitude of a few volts as in W7-X. 
Nevertheless, the much lower temperature (and higher collisionality) in this 
TJ-II sample compared to those considered for the previous two devices, 
leads indeed to conclude that TJ-II exhibits the strongest potential variation
of the three, specially if the electrostatic
energy variation related to $\Phi_1$ is compared to the thermal one. 
A discussion on this comparison for all the cases considered in the present
work is given in section \ref{sec:remarks}. We advance that the electrostatic potential 
variation for the bulk ions reaches up to a 5-8 \% of the thermal energy for the 
low density TJ-II case and up to 10-15 \% for the low density one, but 
despite of this the radial transport when $\Phi_1$ is included
is not as different from the prediction without it could be expected.
This can be observed in figs \ref{fig:tj20_pflux}(left)
and \ref{fig:tj20_pflux}(right) where the comparison of radial particle 
flux density with and without $\Phi_1$ is represented for the high and low density
cases respectively.\\

A possible explanation could rest on the broad spectrum that $\Phi_1$
exhibits in TJ-II, and that could lead to a partial cancellation 
of the effect of Fourier components contributing both to enlarge and diminish the flux level.
In fig. \ref{fig:tj20_phi1maps} (bottom) the main Fourier components are respectively 
represented as a function of $r/a$ for the high density case (left) and 
low density one (right), showing the absence of a clear dominant component.
Although the $\cos\theta$ term is large along the 
radial coordinate in both cases, its amplitude is not much larger than 
other terms similarly large, as e.g the $\sin\theta$, the helical components $\cos(\theta+\phi)$,
$\sin(\theta+\phi)$ or $\sin\theta$.

\section{Final remark on the amplitude of $\Phi_1$}
\label{sec:remarks}

\begin{figure}[t]
\begin{center}
  \includegraphics[width=0.6\textwidth]{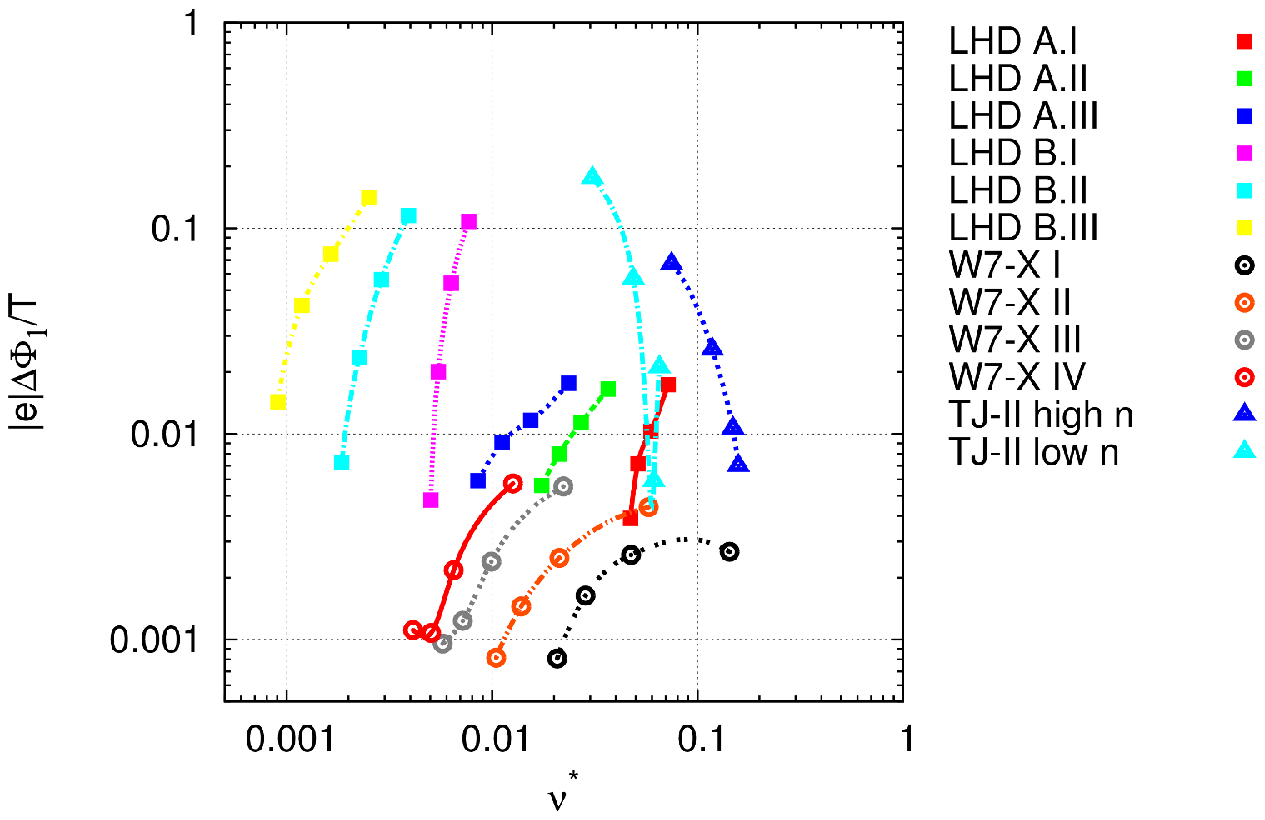}
  \caption{Normalized \textit{peak-to-peak} variation of the potential $\Delta\Phi_{1}$
for all cases considered for LHD, W7-X and TJ-II as a function of the normalized
thermal collision frequency of bulk the ions normalized to the bounce frequency $\omega_b$.}
  \label{fig:dphi1_nu}
\end{center}
\end{figure}

In fig. \ref{fig:dphi1_nu} the maximum variation of electrostatic energy for H$^+$
related to the difference between the maximum and minimum value of the potential on 
the surface $\Delta Phi_1$ is 
represented, as a function of the normalized collision frequency normalized to the bounce frequency, 
$\nu^{*}=\nu_{i,\text{th}}/\omega_{b}$, with $\omega_{b}=v_{\text{th}}\iota/R$.
This ratio $e\Delta\Phi_1/T$ is represented for all the cases considered in the work at
the four radial positions considered in each case. The value 
$e\Delta\Phi_1/T$ provides an estimate of how reliable is the 
standard neoclassical assumption of mono-energetic trajectories for the bulk ions, 
and if multiplied by the corresponding charge state, for the impurity
species of interest.\\
To this respect, having taken C$^{6+}$ as the impurity for the present study,
it is clear that the mono-energetic assumption breaks down in the
LHD and TJ-II cases at the most external position $r/a=0.6$ and $0.8$, that 
correspond to the uppermost points of each curve. This is remarkably
clear at the low collisional LHD set of profiles and the two
cases in TJ-II. In TJ-II the potential variation 
is particularly large considering the much larger collisionality than in 
the LHD set \textit{B}, as it was noted at the end of the previous section. 
At similar collisionality, comparing the LHD set case A.I and TJ-II low density
one, LHD shows roughly an order of magnitude lower $\Phi_1$ than TJ-II. 
Note that in this normalized picture the ratios comparing the strength of $\nabla\Phi_1$ and $\nabla B$
as transport and trapping sources in eqs. \ref{eq:ve1_over_vd} and \ref{eq:as_over_am} 
become of order unit as well.\\
In the figure it can be appreciated how at similar collisional
regime W7-X reaches values of $e\Delta\Phi_{1}/T$ one order of 
magnitude lower that in the LHD high density set, and finds
the usual neoclassical neglect of $\Phi_1$ well valid for the 
parameters considered.

\section{Summary and conclusions}
\label{sec:summary}

The impurity transport is known to be strongly influenced by
the ambipolar part of the electrostatic potential, usually considered as an approximation 
to the full neoclassical electrostatic potential.
In the present work, we have also considered the potential variation
on the flux surface $\Phi_{1}$ after solving the quasi-neutrality equation,
and calculated the impact on the radial particle flux of C$^{6+}$ for the devices
LHD, W7-X and TJ-II, using the code EUTERPE.\\
The calculation have confirmed the initial assumptions on the 
importance of taking $\Phi_{1}$ into account for the transport of impurities, 
given that for these $\Phi_1$ can both modifying substantially the topology of the trapped/passing boundaries or 
driving transport through the related $E\times B$ drift in the same degree than $\nabla B$.\\
In most cases $\Phi_{1}$ has led to appreciable deviations from the 
results framed in the standard neoclassical approach, both mitigating or
enhancing the trend to accumulate.
This was particularly clear in the LHD low collisional cases, 
and in less extent in the higher collisional TJ-II ones, despite the large variation of $\Phi_1$
the latter device exhibit. Finally in W7-X $\Phi_1$
has resulted to be weak enough to be negligible for the equilibrium and 
parameters considered.\\
Although an extension of previous studies, and apart from 
a more exhaustive parameter and magnetic configuration scan, there are still major 
points to address and assumptions to relax in next works: first of all the relaxation of the tracer impurity limit, since  
scenarios with accumulation unavoidably can lead in long discharges to sufficiently high impurity concentration to 
make this approximation break down; once impurities are no longer tracers and must be included
in the quasi-neutrality equation the linear adiabatic response 
considered, specially if they have large charge states, is no longer valid either; a detailed study of how $\Phi_1$ 
and the spacial impurity distribution function couples to each other
could also help to identify what \textit{phase shift} between both can bring mitigated accumulation 
scenarios.

\section{Acknowledgements}
This work was supported by EURATOM and carried out within the framework of the European Fusion Development Agreement. The views and opinions expressed herein do not necessarily reflect those of the European Commision.\\
The calculations were carried out using the HELIOS supercomputer system at Computational Simulation Centre of International Fusion Energy Research Centre (IFERC-CSC), Aomori, Japan, under the Broader Approach collaboration between Euratom and Japan, implemented by Fusion for Energy and JAEA.\\

\label{Bibliography}


\section*{References}

\begin{thebibliography}{10}

\bibitem{Ida_pop_16_056111_2009}
K~Ida \textit{et al.}, {\em Phys. Plasmas} \textbf{16} 056111 (2009).

\bibitem{Yoshinuma_nf_49_062002_2009}
M.~Yoshinuma \textit{et al.}, {\em Nuclear Fusion} \textbf{49} 062002 (2009).

\bibitem{Kallenbach_ppcf_55_124041_2013}
A~Kallenbach \textit{et al.}, {\em Plasma Phys. and Control. Fusion} \textbf{55} 124041 (2013).

\bibitem{Ho_pf_30.2_1987}
D~D.-M Ho and R~M Kulsrud, {\em Phys. Fluids} \textbf{30}(2) 442--461 (1987).

\bibitem{Beidler_nf_51_076001_2011}
C~D Beidler \textit{et al.}, {\em Nuclear Fusion} \textbf{51} 076001 (2011).

\bibitem{Maassberg_ppcf_41_1999}
H~Maa$\ss$berg, C~D Beidler, and E~E Simmet, {\em Plasma Phys. and Control. Fusion} \textbf{41} 1135 (1999).

\bibitem{Mynick_pf_27.8_1984}
H~E Mynick, {\em Physics of Fluids} \textbf{27}(8) 2086 (1984).

\bibitem{Beidler_isw_2005}
C~D Beidler and H~Ma$\ss$berg, {\em 15th International Stellarator Workshop, Madrid} (2005).

\bibitem{Regana_ppcf_55_074008_2013}
J~M Garc\'ia-Rega\~na \textit{et al.}, {\em Plasma Phys. and Control. Fusion} \textbf{55} 074008 (2013).

\bibitem{Kornilov_nf_45.4_2005}
V~Kornilov \textit{et al.}, {\em Nuclear Fusion} \textbf{45}(4) 238 (2005).

\bibitem{Takizuca_jcp_25.3_1977}
T~Takizuka, {\em Journal of Computational Physics} \textbf{25}(3) 205--219 (1977).

\bibitem{Kauffmann_jpcs_260.1_2010}
K~Kauffmann \textit{et al.}, {\em J. Phys.: Conf. Ser.} \textbf{260}(1) 012014 (2010).

\bibitem{Turkin_pop_18_022505_2011}
Y.~Turkin \textit{et al.} {\em Physics of Plasmas} \textbf{18} 022505 (2011).

\bibitem{Hirshman_pf_29_2951_1986}
S~P Hirshman \textit{et al.}, {\em Physics of Fluids} \textbf{29} 2951 (1986).

\bibitem{Rij_pfb_1_563_1989}
W.~I. van Rij and S.~P. Hirshman, {\em Physics of Fluids B} \textbf{1} 563 (1989).

\bibitem{Beidler_ppcf_43_2001}
C~D Beidler and H~Ma$\ss$berg, {\em Plasma Phys. and Control. Fusion} \textbf{43} 1131--1148 (2001).

\bibitem{Viezzer_ppcf_55_124037_2013}
E~Viezzer \textit{et al.} {\em Plasma Phys. and Control. Fusion} \textbf{55} 124037 (2013).

\bibitem{Arevalo_nf_54_013008_2014}
J~Ar\'evalo \textit{et al.}, {\em Nuclear Fusion} \textbf{54} 013008 (2014).

\bibitem{Pedrosa_arXiv_1404.0932_2014}
M~A Pedrosa \textit{et al.}, {\em submitted to Nuclear Fusion} (2014), preprint: http://arxiv.org/abs/1404.0932.

\bibitem{Garren_pfb_3_2822_1991}
D~A Garren and A~H Boozer, {\em Physics of Fluids B} \textbf{3} 2822 (1991).

\end{thebibliography}


\end{document}